\documentclass[aps,prd,amsmath,amssymb,showpacs,10pt,superscriptaddress]{revtex4-2}
\usepackage{graphicx}
\usepackage{color}
\usepackage{slashed}
\usepackage{txfonts}
\usepackage{subfigure}
\usepackage{enumerate,multirow}
\usepackage{physics}
\usepackage[font=footnotesize]{caption}
\usepackage[breaklinks,colorlinks,urlcolor=blue,linkcolor=red,anchorcolor=magenta,citecolor=blue]{hyperref}
\usepackage{cleveref}
\usepackage{orcidlink}
\crefname{table}{Table}{Tables}
\crefname{equation}{Eq.}{Eqs.}
\crefname{figure}{Fig.}{Figs.}
\crefname{section}{Sec.}{Secs.}
\allowdisplaybreaks
\newcommand{\ihep}{\affiliation{Institute of High Energy Physics, Chinese Academy of Sciences, Beijing 100049, People's Republic of China}}
\newcommand{\ucas}{\affiliation{University of Chinese Academy of Sciences, Beijing 100049, People's Republic of China}}
\newcommand{\tju}{\affiliation{Center for Joint Quantum Studies and Department of Physics, School of Science, Tianjin University, Tianjin 300350, China}}
\begin{document}
\title{Triangle singularity with different dynamical couplings}
\author{Jun Wang\orcidlink{0000-0002-0508-3349}}\email{junwang@ihep.ac.cn}
\ihep\ucas
\author{Xiao-Hai Liu\orcidlink{}}\email{xiaohai.liu@tju.edu.cn}
\tju
\author{Qiang Zhao\orcidlink{0000-0001-8334-2580}}\email{zhaoq@ihep.ac.cn}
\ihep\ucas
\begin{abstract}
  We carry out a systematic study of the impact of dynamical couplings of the vertices on the manifestations of the triangle singularity (TS) in the loop amplitude.
  Twelve combinations of external and internal spin-parity quantum numbers, covering the $PPS$, $VPP$, $VVP$, $VPA$, $VVV$, $VVA$, and $VVS$ effective vertices, are enumerated. Each loop amplitude is reduced via the Passarino--Veltman decomposition into scalar integrals $A_{0}$, $B_{0}$, and $C_{0}$, whose Lorentz-scalar coefficients encode the dynamical content of the vertices. Using the masses and kinematics of the $\eta(1405)\to\pi^{+}\pi^{-}\pi^{0}$ transition via the $K^{*}\bar{K}K$ loop with the intermediate $f_{0}(980)$ resonance in the $\pi^+\pi^-$ invariant spectrum as a common numerical benchmark for the twelve spin-parity assignments, we find that the same loop topology produces three distinct phenomena which manifest through the lineshapes, i.e. a narrow peak dominated by the $C_{0}$ contribution; a suppressed peak in which the $B_{0}$ and $C_{0}$ function contributions are comparable; and a cusp-like profile in which the coefficient of the leading Landau logarithm vanishes at the physical TS point. These results indicate that the visibility of a TS enhancement is not fixed by the Landau condition alone but reshaped by the detailed dynamics involved in the loop transitions.
\end{abstract}

\maketitle

\section{Introduction}
The triangle singularity (TS), also known as Landau singularity, is a leading singularity of a one-loop triangle diagram. It can occur when the three internal particles simultaneously approach their mass shells in a pinched route and propagate along a classically allowed configuration~\cite{Landau:1959fi,Coleman:1965xm,Cutkosky:1960sp}. Once the corresponding kinematic condition is satisfied, the loop integral can develop a logarithmic enhancement that may appear as a narrow peak in the invariant-mass spectrum of the two final-state particles, thereby mimicking the lineshape of a genuine resonance~\cite{Wu:2011yx,Wu:2012pg,Wang:2013cya,Wang:2013hga,Liu:2015taa,Guo:2019twa}. In the decay processes of hadrons there exist abundant open-channel thresholds, providing many kinematic windows for such configurations. Nowadays, more and more effects from the TS mechanism are investigated in theory and identified in experiment, which provide a special way for understanding the near-threshold dynamics in hadron-hadron scatterings~\cite{Wu:2011yx,Wang:2013cya,Wang:2013hga,Wang:2011yh,Guo:2019twa,Liu:2017rhx,Liu:2015taa,Szczepaniak:2015eza,Dong:2020hxe}.

Over the past decade, the TS mechanism has been invoked in discussions of a broad range of experimental phenomena. The anomalously enhanced isospin-violating effect in the $\eta(1405/1475)\to 3\pi$ decay observed by BESIII has been attributed to the combined effect of a $K^{*}\bar{K}K$ triangle loop and $f_{0}(980)$--$a_{0}(980)$ mixing~\cite{Wu:2011yx,Aceti:2012dj,Achasov:2015uua,Du:2019idk}. The charged charmoniumlike state $Z_{c}(3900)$ observed in $Y(4260)\to\pi^{+}\pi^{-}J/\psi$ has also been examined from the perspective of a $DD_1\bar{D}^{*}$ triangle mechanism~\cite{Wang:2013cya,Wang:2013hga}, and similar studies have been extended to a number of $XYZ$ candidates~\cite{Wang:2011yh,Wang:2013cya,Wang:2013hga,Liu:2013vfa,Li:2013zcr,Liu:2014spa,Guo:2014iya,Liu:2015cah,Cao:2017lui,Yu:2024sqv,Zhang:2025fcv,Wang:2025zbv}. The hidden-charm pentaquark states $P_{c}$ reported by LHCb in $\Lambda_{b}\to J/\psi p K$ decays have likewise been examined within a triangle-mechanism interpretation~\cite{Guo:2015umn,Liu:2015fea,Bayar:2016ftu,Guo:2016bkl,Liu:2016dli,Nakamura:2021qvy}. A direct evidence for the TS mechanism is in connection with the axial-vector candidate $a_{1}(1420)$~\cite{Liu:2015taa,Mikhasenko:2015oxp}. Meanwhile, it shows that experimental data for the $N(1700)$ region~\cite{Roca:2017bvy}, the $X(5568)$~\cite{Liu:2016xly}, the Dalitz plot of $\Lambda_{c}^{+}\to pK^{-}\pi^{+}$~\cite{Liu:2019dqc}, the $X_{0,1}(2900)$~\cite{Liu:2020orv}, etc.  In most of these studies, it is generally emphasized that the satisfaction of the Landau condition can produce narrow peaking enhancement.
But it should be recognized that in all these different cases the significance of the TS structure would also be affected by the dynamical details for the vertices.
So far, systematic studies of the dynamical coupling effects with the TS mechanism have not yet been explored. This motivates us to investigate the manifestations of the TS mechanism when different coupling vertices are present.


For a fixed Coleman--Norton kinematic configuration~\cite{Coleman:1965xm}, the detailed structure of the three vertices—including the partial wave carried by each coupling ($S$ wave or $P$ wave), the tensor structures associated with the vector or axial-vector intermediate states, and the Lorentz scalars generated in the amplitude decomposition, determines the relative weights of the three-point scalar integral $C_{0}$ and the two-point scalar integrals $B_{0}$. When $C_{0}$ dominates, the invariant-mass spectrum generally exhibits a narrow TS peak. Comparable $B_{0}$ contributions can interfere with the triangle term and reduce the prominence of the peak. If the leading Landau logarithm is suppressed by the vertex structure, the remaining ordinary-threshold dependence and smooth background of the coherent amplitude can instead produce a cusp-like profile, without requiring an explicit $B_{0}$ term to dominate. To the best of our knowledge, a systematic comparison of how different dynamical couplings modify the TS lineshape for a fixed loop topology is still lacking. It is useful and necessary to investigate the diversity of the TS lineshapes  from this perspective.

In this work, we systematically investigate the effect of the dynamical couplings at the triangle vertices on the TS lineshape. We consider ground-state $0^{-}$ or $1^{-}$ intermediate particles and allow the three vertices to couple through either an $S$ or $P$ wave, leading to twelve combinations of external and internal spin-parity quantum numbers. The corresponding effective Lagrangians cover seven types of vertices, namely $PPS$, $VPP$, $VVP$, $VPA$, $VVV$, $VVA$, and $VVS$~\cite{Cao:2023gfv,Cheng:2021nal,Qiu:2022ktc,Dulat:2005in}. With the aid of FeynCalc~\cite{Mertig:1990an,Shtabovenko:2016sxi,Shtabovenko:2020gxv,Shtabovenko:2023idz}, we use the Passarino--Veltman reduction~\cite{Passarino:1978jh} to express each loop amplitude as a linear combination of the scalar integrals $A_{0}$, $B_{0}$, and $C_{0}$. The Lorentz-scalar coefficients multiplying these integrals encode the dynamical information of the vertices under consideration. Comparing the relative sizes of these contributions near the kinematic position determined by the Landau equation allows us to identify the role of each scalar integral in shaping the spectrum. For the numerical comparison, we adopt the masses and kinematics of the $\eta(1405)\to\pi^{+}\pi^{-}\pi^{0}$ system with a $K^{*}\bar{K}K$ loop as a common benchmark for the twelve spin-parity assignments, using the $f_{0}(980)$ parameters for the final-state resonance factor. This is the kinematic region where $f_1(1420)$ and $h_1(1380/1423)$ can have an $S$-wave coupling to the $K^*\bar{K}$ channel. We find that, even for the same loop topology and kinematic condition, different vertex structures can generate three qualitatively distinct lineshapes.


The remainder of this paper is organized as follows. The formalism is introduced in \cref{sec:formalism}; the numerical results and their discussion are presented in \cref{sec:results}; and the main conclusions are summarized in \cref{sec:summary}.

\section{Formalism}\label{sec:formalism}

\begin{figure}[!ht]
  \centering
  \includegraphics[width=0.5\textwidth]{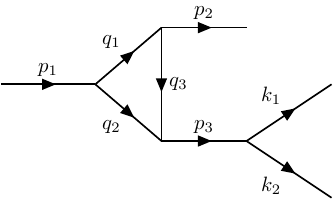}
  \caption{Triangle loop diagram}\label{fig:TS}
\end{figure}
The contribution of the triangle loop diagram for the $1\to 3$ process is shown in \cref{fig:TS}.
For a scalar triangle loop diagram, the amplitude is given by
\begin{equation}
  i \mathcal{M}=\int \frac{d^{4}q_{3}}{(2\pi)^{4}}\frac{1}{(q_1^{2}-m_1^{2})(q_2^{2}-m_2^{2})(q_3^{2}-m_3^{2})}\,,\quad q_1=p_2+q_3,q_2=p_3-q_3.
\end{equation}
Using the Feynman parameterization method and denoting the shifted loop momentum by $q$, this can be expressed as
\begin{equation}
  i \mathcal{M}= \int \frac{d^{4}q}{(2\pi)^{4}} \iiint dx_{1}dx_{2}dx_{3} \frac{2\delta(\sum_{i=1}^{3}x_{i}-1)}{(q^{2}-D)^{3}}\,, \quad D=Y_{ij}x_{i}x_{j},\quad Y_{ij}=\frac{1}{2}[m_{i}^{2}+m^{2}_{j}-\Delta_{q_{ij}}]\,.
\end{equation}
where
\begin{equation}
  \Delta_{q_{ij}}=\Delta_{q_{ji}},\quad \Delta_{q_{12}}=p_1^{2},\quad \Delta_{q_{23}}=p_3^{2},\quad \Delta_{q_{31}}=p_2^{2}\,.
\end{equation}
The location of the triangle singularity is determined by the Landau equations~\cite{Landau:1959fi,Du:2019idk,Liu:2017rhx}. Denoting $s_1=p_1^{2}$, $s_2=p_2^{2}$, and $s_3=p_3^{2}$, with $s_1$ and $s_2$ fixed, $s_3$ satisfies the following equation:
\begin{equation}
  \det (Y_{ij})=0\implies
  \begin{aligned}
    \frac{1}{4} \{m_1^2 [m_2^2 (-s_1+s_2+s_3)+m_3^2 (s_1-s_2+s_3)+(s_1+s_2-s_3) s_3]-s_3 m_1^4-s_2m_2^4 \\-s_1 [m_3^2 (s_1-s_2-s_3)+m_3^4+s_2 s_3]+m_2^2 [m_3^2 (s_1+s_2-s_3)+s_2 (s_1-s_2+s_3)]\}=0\,. \label{eq:laeq}
  \end{aligned}
\end{equation}
The equation for $s_{3}$ has two solutions. In the kinematic window considered here, the physical TS corresponds to the branch $s_{3}^{-}$ that satisfies the Coleman--Norton condition, rather than merely lying within the kinematically allowed phase space:
\begin{equation}
  s_{3}=s_{3}^{-}=(m_2+m_3)^{2}+\frac{1}{2m_1^{2}}[(m_1^{2}+m_3^{2}-s_2)(s_1-m_1^{2}-m_2^{2})-4m_1^{2}m_2m_3-\sqrt{\lambda[s_1,m_1^{2},m_2^{2}]\lambda[s_2,m_1^{2},m_3^{2}]}]\,.\label{eq:s3ts}
\end{equation}
where $\lambda(x,y,z)=x^2+y^2+z^2-2xy-2yz-2zx$ is the Källén triangle function~\cite{Kallen:1964lxa}.

Although the position of the triangle singularity is fixed, different dynamical couplings can significantly affect the magnitude of the triangle singularity and even cancel its contribution. To investigate the impact of different dynamical couplings on the triangle singularity, we first consider the intermediate particles as ground-state $S$-wave particles with quantum numbers $J^{P}=0^{-},1^{-}$. For the quantum numbers of the initial and final states, we consider $S$-wave or $P$-wave couplings of ground-state particles, leading to the following possibilities.
The kinematic relations are
\begin{equation}
  p_1=q_1+q_2,\quad q_1=q_{3}+p_2,\quad p_3=q_2+q_3
\end{equation}
where $p_{1,2,3}$ denote the initial and final state particles, $q_{1,2,3}$ denote the intermediate particles, and $V_{1,2,3}$ represent the partial wave types of the vertex couplings $p_1q_1q_2$, $p_2q_1q_3$, and $p_3q_2q_3$, respectively. For the process $p_{3}\to k_1 +k_2$, which depends only on the external momenta, we focus on the cases $1^{+}\to 1^{-}+0^{-}$, $0^{+}\to 0^{-}+0^{-}$, $1^{-}\to 0^{-}+0^{-}$, and $0^{-}\to 1^{-}+0^{-}$.
\begin{table}[!ht]
  \centering
  \caption{Possible quantum number combinations for the triangle diagram processes}
  \begin{ruledtabular}    \label{tab:possible_quantum_numbers}
    \begin{tabular}{cccccccccccc}
      Serial Number&$p_{1}$&$p_{2}$&$p_{3}$&$q_1$&$q_{2}$&$q_{3}$&$k_1$&$k_{2}$&$V_1$&$V_2$&$V_{3}$\\\hline
      1&$1^{+}$&$0^{-}$&$0^{+}$&$1^{-}$&$0^{-}$&$0^{-}$&$0^{-}$&$0^{-}$&$S$&$P$&$S$\\
      2&$1^{+}$&$0^{-}$&$1^{-}$&$1^{-}$&$0^{-}$&$0^{-}$&$0^{-}$&$0^{-}$&$S$&$P$&$P$\\
      3&$1^{+}$&$0^{-}$&$1^{-}$&$1^{-}$&$1^{-}$&$0^{-}$&$0^{-}$&$0^{-}$&$S$&$P$&$P$\\
      4&$1^{+}$&$0^{-}$&$1^{+}$&$1^{-}$&$1^{-}$&$0^{-}$&$1^{-}$&$0^{-}$&$S$&$P$&$S$\\
      5&$1^{-}$&$0^{-}$&$0^{-}$&$1^{-}$&$1^{-}$&$0^{-}$&$1^{-}$&$0^{-}$&$P$&$P$&$P$\\
      6&$1^{-}$&$0^{-}$&$1^{-}$&$1^{-}$&$0^{-}$&$0^{-}$&$0^{-}$&$0^{-}$&$P$&$P$&$P$\\
      7&$1^{-}$&$0^{-}$&$1^{-}$&$1^{-}$&$1^{-}$&$0^{-}$&$0^{-}$&$0^{-}$&$P$&$P$&$P$\\
      8&$1^{-}$&$0^{-}$&$1^{+}$&$1^{-}$&$1^{-}$&$0^{-}$&$1^{-}$&$0^{-}$&$P$&$P$&$S$\\
      9&$0^{+}$&$0^{-}$&$1^{+}$&$1^{-}$&$1^{-}$&$0^{-}$&$1^{-}$&$0^{-}$&$S$&$P$&$S$\\
      10&$0^{-}$&$0^{-}$&$0^{+}$&$1^{-}$&$0^{-}$&$0^{-}$&$0^{-}$&$0^{-}$&$P$&$P$&$S$\\
      11&$0^{-}$&$0^{-}$&$1^{-}$&$1^{-}$&$0^{-}$&$0^{-}$&$0^{-}$&$0^{-}$&$P$&$P$&$P$\\
      12&$0^{-}$&$0^{-}$&$1^{-}$&$1^{-}$&$1^{-}$&$0^{-}$&$0^{-}$&$0^{-}$&$P$&$P$&$P$
    \end{tabular}
  \end{ruledtabular}
\end{table}

We consider four types of particles: $S$ (scalar), $P$ (pseudoscalar), $V$ (vector), and $A$ (axial-vector). The amplitudes are labeled by their serial numbers. The coupling types involved are $PPS$, $VPP$, $VVP$, $VPA$, $VVV$, $VVA$, and $VVS$. The corresponding effective Lagrangians are given in Refs.~\cite{Cao:2023gfv,Cheng:2021nal,Qiu:2022ktc,Dulat:2005in}:
\begin{equation}
  \begin{aligned}
    \mathcal{L}_{PPS}=&g_{PPS}P_{1}P_{2}S\,,\quad \mathcal{L}_{VPP}=ig_{VPP}(P_{1}\partial_\mu P_{2}-\partial_{\mu}P_{1}P_{2})V^\mu,\quad  \mathcal{L}_{VPA}=g_{VPA}V_{\mu}A^{\mu}P\,, \\
    \mathcal{L}_{VVP}=&g_{VVP}\varepsilon_{\alpha \beta \mu \nu}\partial^{\alpha}V^{\mu}_{1}\partial^{\beta}V_{2}^{\nu}P\,,\quad\mathcal{L}_{VVA}=i g_{VVA}\varepsilon_{\alpha \beta \mu \nu}\partial^{\alpha}A^{\beta}V_{1}^{\mu}V_{2}^{\nu}\,,\\
    \mathcal{L}_{VVV}=&ig_{VVV}(V_{1,\nu}\overleftrightarrow{\partial^{\mu}}V_2^{\nu}V_{3,\mu}+V_{1,\mu}V_2^{\nu}\overleftrightarrow{\partial^{\mu}}V_{3,\nu}+V_{2,\mu}V_3^{\nu}\overleftrightarrow{\partial^{\mu}}V_{1,\nu})\,,\quad \mathcal{L}_{VVS}=g_{VVS}V_{1}^{\mu}V_{2\mu}S\,.
  \end{aligned}
\end{equation}

The masses of the intermediate particles in the loop diagram are denoted by $m_{1,2,3}$, and the masses of the external particles $p_{1,2,3}$ by $M_{1,2,3}$. The invariant mass squared of the final state is $s_3=p_{3}^{2}=(k_1+k_2)^2$. The amplitudes can be written as
\begin{align}
  i \mathcal{M}_{1}=&g_{1}\int \frac{d^{4}q_{3}}{(2\pi)^{4}}\frac{\varepsilon_{1i}^{\mu}g_{\mu\nu}(q_{3}^{\nu}-p_{2}^{\nu})}{(q_1^{2}-m_1^{2})(q_2^{2}-m_2^{2})(q_3^{2}-m_3^{2})}\frac{1}{s_3-M^{2}_{3}+i M_{3}\Gamma_{3}}\ ,\\
  i \mathcal{M}_{2}=&g_{2}\int \frac{d^{4}q_{3}}{(2\pi)^{4}}\frac{\varepsilon_{1i}^{\mu}g_{\mu\nu}(p_{2}-q_{3})^{\nu}(q_{2}-q_{3})^{\alpha}}{(q_1^{2}-m_1^{2})(q_2^{2}-m_2^{2})(q_3^{2}-m_3^{2})}\frac{g_{\alpha\beta}(k_{1}-k_2)^{\beta}}{s_3-M^{2}_{3}+i M_{3}\Gamma_{3}}\ ,\\
  i \mathcal{M}_{3}=&g_{3}\int \frac{d^{4}q_{3}}{(2\pi)^{4}}\frac{\varepsilon_{\mu\nu \rho \sigma}\varepsilon_{1i}^{\mu}p_{1}^{\nu}(p_{2}-q_{3})^{\rho}g^{\sigma \delta}\varepsilon_{\lambda \tau\alpha \delta}q_{2}^{\lambda}p_{3}^{\tau}}{(q_1^{2}-m_1^{2})(q_2^{2}-m_2^{2})(q_3^{2}-m_3^{2})}\frac{g^{\alpha\beta}(k_{1}-k_2)_{\beta}}{s_3-M^{2}_{3}+i M_{3}\Gamma_{3}}\ ,\\
  i \mathcal{M}_{4}=&g_{4}\int \frac{d^{4}q_{3}}{(2\pi)^{4}}\frac{\varepsilon_{\mu\nu \rho \sigma}\varepsilon_{1i}^{\mu}p_{1}^{\nu}(p_{2}-q_{3})^{\rho}}{(q_1^{2}-m_1^{2})(q_2^{2}-m_2^{2})(q_3^{2}-m_3^{2})}\frac{\varepsilon_{1f}^{*\sigma}}{s_3-M^{2}_{3}+i M_{3}\Gamma_{3}}\ ,\\
  i \mathcal{M}_{5}=&g_{5}\int \frac{d^{4}q_{3}}{(2\pi)^{4}}\frac{\varepsilon_{1i}^{\mu}(g_{\rho\sigma}(q_{2}-q_{1})_{\mu}-g_{\nu \rho}g_{\mu\sigma}(p_1+q_2)^{\nu}+g_{\mu\rho}g_{\alpha \sigma}(q_{1}+p_1)^{\alpha})}{(q_1^{2}-m_1^{2})(q_2^{2}-m_2^{2})(q_3^{2}-m_3^{2})}\\\nonumber&\times \frac{(p_2-q_3)^{\rho}(p_3+q_3)^{\sigma}\varepsilon_{1f}^{*}\cdot (k_2+p_3)}{s_3-M^{2}_{3}+i M_{3}\Gamma_{3}}\ ,\\
  i \mathcal{M}_{6}=&g_{6}\int \frac{d^{4}q_{3}}{(2\pi)^{4}}\frac{\varepsilon_{\mu \nu \rho\sigma}\varepsilon_{1i}^{\mu}p_{1}^{\nu}q_{1}^{\sigma}g^{\rho \lambda}(p_2-q_3)_{\lambda}(q_2-q_3)^{\tau}}{(q_1^{2}-m_1^{2})(q_2^{2}-m_2^{2})(q_3^{2}-m_3^{2})}\frac{(k_1-k_2)_{\tau}}{s_3-M^{2}_{3}+i M_{3}\Gamma_{3}}\ ,\\
  i \mathcal{M}_{7}=&g_{7}\int \frac{d^{4}q_{3}}{(2\pi)^{4}}\frac{\varepsilon_{1i}^{\mu}(g_{\rho\sigma}(q_{2}-q_{1})_{\mu}-g_{\nu \rho}g_{\mu\sigma}(p_1+q_2)^{\nu}+g_{\mu\rho}g_{\alpha \sigma}(q_{1}+p_1)^{\alpha})}{(q_1^{2}-m_1^{2})(q_2^{2}-m_2^{2})(q_3^{2}-m_3^{2})}\\\nonumber&\times \frac{(p_2-q_3)^{\rho}g^{\sigma \eta}\varepsilon_{\eta \delta \lambda \tau}q_{2}^{\delta}p_3^{\tau}(k_1-k_2)^{\lambda}}{s_3-M^{2}_{3}+i M_{3}\Gamma_{3}}\ ,\\
  i \mathcal{M}_{8}=&g_{8}\int \frac{d^{4}q_{3}}{(2\pi)^{4}}\frac{\varepsilon_{1i}^{\mu}(g_{\rho\sigma}(q_{2}-q_{1})_{\mu}-g_{\nu \rho}g_{\mu\sigma}(p_1+q_2)^{\nu}+g_{\mu\rho}g_{\alpha \sigma}(q_{1}+p_1)^{\alpha})}{(q_1^{2}-m_1^{2})(q_2^{2}-m_2^{2})(q_3^{2}-m_3^{2})}\\\nonumber&\times \frac{(p_2-q_3)^{\rho}\varepsilon_{1f}^{*\sigma}}{s_3-M^{2}_{3}+i M_{3}\Gamma_{3}}\ ,\\
  i \mathcal{M}_{9}=&g_{9}\int \frac{d^{4}q_{3}}{(2\pi)^{4}}\frac{(p_2-q_3)^{\mu}\varepsilon_{1f\mu}^{*}}{(q_1^{2}-m_1^{2})(q_2^{2}-m_2^{2})(q_3^{2}-m_3^{2})} \frac{1}{s_3-M^{2}_{3}+i M_{3}\Gamma_{3}}\ ,\\
  i \mathcal{M}_{10}=&g_{10}\int \frac{d^{4}q_{3}}{(2\pi)^{4}}\frac{(p_2-q_3)^{\mu}(p_1+q_2)_{\mu}}{(q_1^{2}-m_1^{2})(q_2^{2}-m_2^{2})(q_3^{2}-m_3^{2})} \frac{1}{s_3-M^{2}_{3}+i M_{3}\Gamma_{3}}\ ,\\
  i \mathcal{M}_{11}=&g_{11}\int \frac{d^{4}q_{3}}{(2\pi)^{4}}\frac{(p_{1}+q_{2})^{\mu}(p_2-q_3)_{\mu}(q_3-q_2)^{\nu}}{(q_1^{2}-m_1^{2})(q_2^{2}-m_2^{2})(q_3^{2}-m_3^{2})} \frac{(k_1-k_2)_{\nu}}{s_3-M^{2}_{3}+i M_{3}\Gamma_{3}}\ ,\\
  i \mathcal{M}_{12}=&g_{12}\int \frac{d^{4}q_{3}}{(2\pi)^{4}}\frac{\varepsilon_{\mu\nu \rho\sigma}q_{1}^{\nu}q_{2}^{\sigma}(p_2-q_3)^{\mu}g^{\rho \alpha}\varepsilon_{\alpha \lambda \delta \tau}q_{2}^{\lambda}}{(q_1^{2}-m_1^{2})(q_2^{2}-m_2^{2})(q_3^{2}-m_3^{2})} \frac{p_{3}^{\tau}(k_1-k_2)^{\delta}}{s_3-M^{2}_{3}+i M_{3}\Gamma_{3}}\ .
\end{align}
Here, $g_{i}$ ($i=1,2,\cdots,12$) denotes the product of coupling constants involved in the loop diagram. $\varepsilon_{1i}$ is the polarization vector of the initial-state particle, and $\varepsilon_{1f}^{*}$ is the polarization vector of the final-state particle.

The one-loop amplitudes can be decomposed into scalar loop integrals with Lorentz-scalar coefficients~\cite{Passarino:1978jh,Du:2019idk}.
\begin{align}
  i \mathcal{M}_{i}/g_{i}=\qty(\sum_{j}A_{i}^{j} \mathbf{A}_{0j}+\sum_{j}B_{i}^{j} \mathbf{B}_{0j}+C_{i}\mathbf{C}_{0})\frac{1}{s_{3}-M^{2}_{3}+i M_{3}\Gamma_{3}}\,,
\end{align}
The scalar loop integrals $\mathbf{A}_{0j}$, $\mathbf{B}_{0j}$, and $\mathbf{C}_{0}$ are defined as
\begin{equation}
  \begin{aligned}
    \mathbf{A}_{01}=&\int \frac{d^{4}q_{3}}{(2\pi)^{4}}\frac{1}{D_{1}}\,,\quad \mathbf{A}_{02}=\int \frac{d^{4}q_{3}}{(2\pi)^{4}}\frac{1}{D_{2}}\,,\quad \mathbf{A}_{03}=\int \frac{d^{4}q_{3}}{(2\pi)^{4}}\frac{1}{D_{3}}\,,\quad\\
    \mathbf{B}_{01}=&\int \frac{d^{4}q_{3}}{(2\pi)^{4}}\frac{1}{D_{2}D_{3}}\,,\quad
    \mathbf{B}_{02}=\int \frac{d^{4}q_{3}}{(2\pi)^{4}}\frac{1}{D_{1}D_{3}}\,,\quad \mathbf{B}_{03}=\int \frac{d^{4}q_{3}}{(2\pi)^{4}}\frac{1}{D_{1}D_{2}}\,,\quad\\
    \mathbf{C}_{0}=&\int \frac{d^{4}q_{3}}{(2\pi)^{4}}\frac{1}{D_{1}D_{2}D_{3}}\,,\\
  \end{aligned}
\end{equation}
where $D_{1}=q_1^{2}-m_1^{2}$, $D_2=q_2^{2}-m_2^{2}$, and $D_{3}=q_3^{2}-m_3^{2}$.
The explicit expressions for $A_{i}^{j}$, $B_{i}^{j}$, and $C_{i}$ can be found in Appendix \ref{appendix:A} (see~\cref{tab:coeff1,tab:coeff2,tab:coeff3,tab:coeff4,tab:coeff5,tab:coeff6,tab:coeff7,tab:coeff8,tab:coeff9,tab:coeff10,tab:coeff11,tab:coeff12}).
We also consider the case without an intermediate resonance in the final state, i.e., $q_2 q_3 \to k_1 k_2$ proceeds via a contact interaction. In this case, the corresponding amplitude is obtained by removing the Breit-Wigner factor, and the amplitude is given by
\begin{equation}
  i \mathcal{M}_{i}^{\text{C}}= i \mathcal{M}_{i}\times  (s_{3}-M_{3}^{2}+i M_{3}\Gamma_{3})\,.
\end{equation}
To regulate the ultraviolet behavior of the loop integrals, we introduce the form factor
\begin{equation}
  \mathcal{F}(q^{2})=\prod_{i=1}^{3}\frac{\Lambda_{i}^{2}-m_{i}^{2}}{\Lambda_{i}^{2}-q_{i}^{2}}\,,
\end{equation}
where $\Lambda_{i}=m_{i}+\alpha \Lambda_{\text{QCD}}$, with $\Lambda_{\text{QCD}}=0.22~\mathrm{GeV}$ denoting the QCD energy scale and $\alpha$ the cutoff parameter. For small values of the cutoff parameter, the form factor may introduce additional singularities. We therefore choose $\alpha=2$, so that the singularities induced by the form factor remain far from the energy region of interest. For brevity, the form factor $\mathcal{F}(q^{2})$ is not written explicitly in the amplitude expressions above, but it is included in the complete amplitudes used to obtain the spectra and the relative contributions in \cref{tab:relativeloop}. The coefficient scans and the separate scalar-integral plots shown below instead use the physical masses without the form-factor subtraction, as specified in the corresponding captions.
The form factor can be implemented by mass subtraction, since
\begin{equation}
  \frac{1}{q_r^{2}-m_r^{2}}\frac{\Lambda_r^{2}-m_r^{2}}{\Lambda_r^{2}-q_r^{2}}=\frac{1}{q_r^{2}-m_r^{2}}-\frac{1}{q_r^{2}-\Lambda_r^{2}}\,.
\end{equation}
We therefore define the eight-term subtraction operator
\begin{equation}
  \mathcal{R}[X]=(1-\mathsf{T}_{1})(1-\mathsf{T}_{2})(1-\mathsf{T}_{3})X,\qquad \mathsf{T}_{r}X=\left.X\right|_{m_r^{2}\to\Lambda_r^{2}}\,.\label{eq:monopole-subtraction}
\end{equation}
Here each replacement acts on the corresponding propagator-mass parameter, with the external kinematics, couplings, and cutoff values held fixed. After the scalar-integral decomposition, $\mathcal{R}$ must act on each complete coefficient-times-integral product, rather than on the scalar integral alone. The three propagator masses are kept independent during the subtraction; physical mass equalities are imposed only afterwards. The analytic coefficients in Appendix~\ref{appendix:A} are those of the unregulated amplitudes.
\section{Numerical Results and Discussions}\label{sec:results}
\subsection{Spectra and scalar-integral decomposition}
\begin{figure}[!ht]
  \centering
  \subfigure[]{%
    \includegraphics[width=0.33\textwidth]{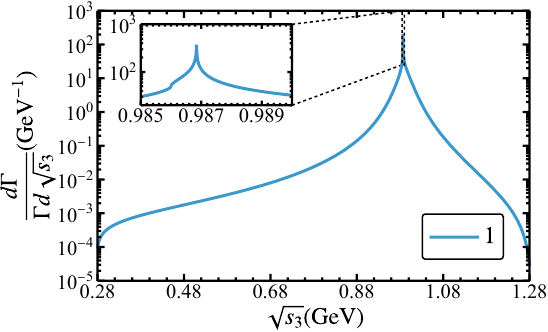}\label{fig:2-1}%
  }%
  \subfigure[]{%
    \includegraphics[width=0.33\textwidth]{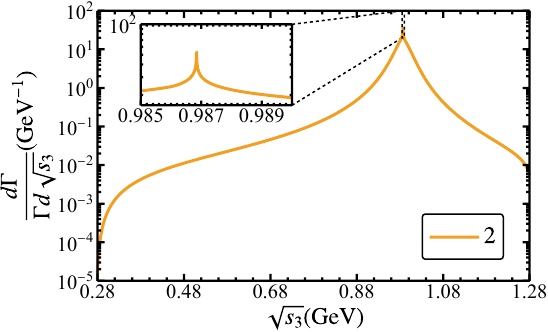}\label{fig:2-2}%
  }%
  \subfigure[]{%
    \includegraphics[width=0.33\textwidth]{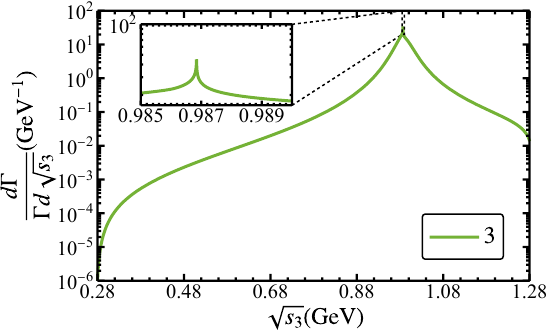}\label{fig:2-3}%
  }
  \subfigure[]{%
    \includegraphics[width=0.33\textwidth]{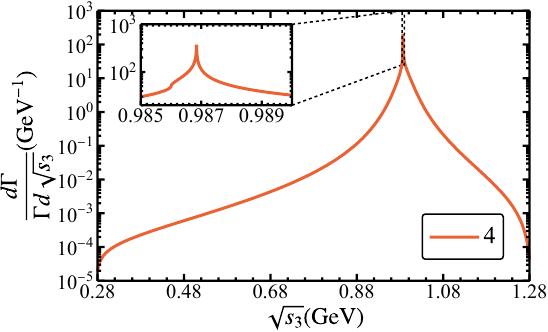}\label{fig:2-4}%
  }%
  \subfigure[]{%
    \includegraphics[width=0.33\textwidth]{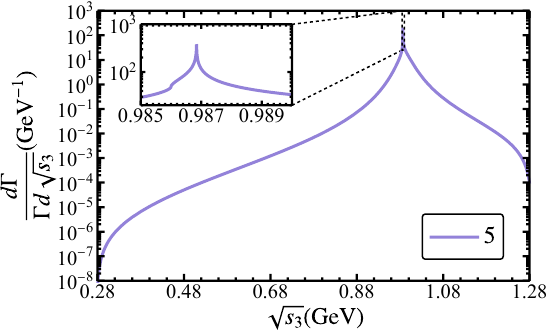}\label{fig:2-5}%
  }%
  \subfigure[]{%
    \includegraphics[width=0.33\textwidth]{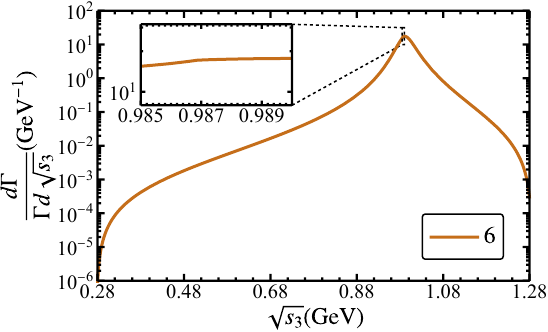}\label{fig:2-6}%
  }
  \subfigure[]{%
    \includegraphics[width=0.33\textwidth]{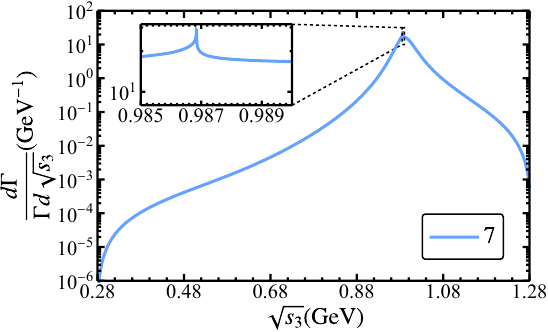}\label{fig:2-7}%
  }%
  \subfigure[]{%
    \includegraphics[width=0.33\textwidth]{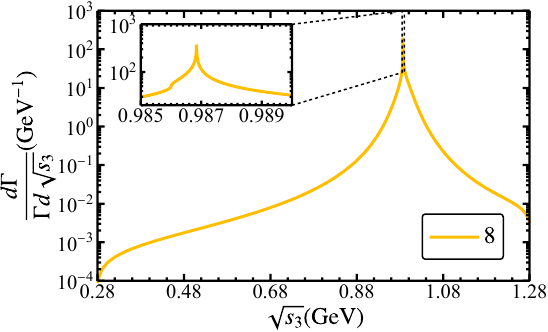}\label{fig:2-8}%
  }%
  \subfigure[]{%
    \includegraphics[width=0.33\textwidth]{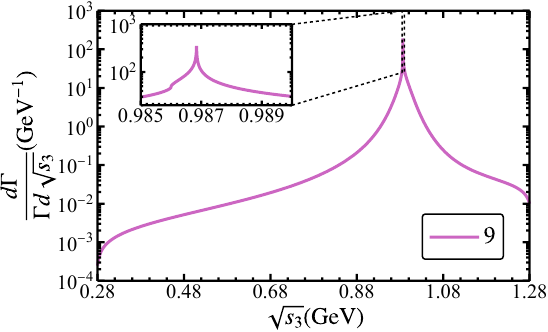}\label{fig:2-9}%
  }
  \subfigure[]{%
    \includegraphics[width=0.33\textwidth]{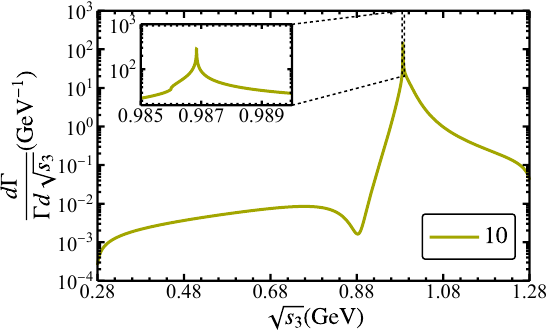}\label{fig:2-10}%
  }%
  \subfigure[]{%
    \includegraphics[width=0.33\textwidth]{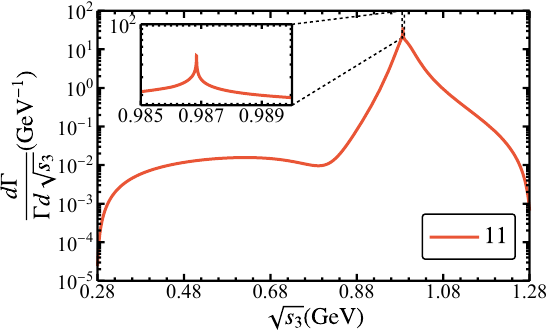}\label{fig:2-11}%
  }%
  \subfigure[]{%
    \includegraphics[width=0.33\textwidth]{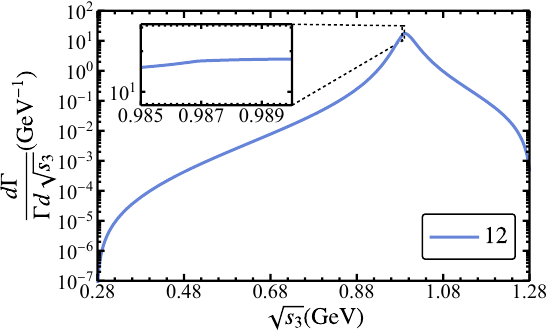}\label{fig:2-12}%
  }%
  \caption{Invariant-mass spectra $\sqrt{s_3}=M(k_1k_2)$ for the twelve spin-parity assignments in \cref{tab:possible_quantum_numbers}, using the common mass inputs and including the final-state resonance factor with the $f_{0}(980)$ parameters.}
  \label{fig:spectrum1}
\end{figure}

\begin{figure}[!ht]
  \centering
  \subfigure[]{%
    \includegraphics[width=0.33\textwidth]{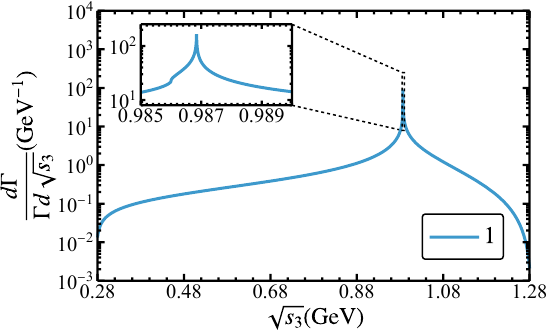}\label{fig:3-1}%
  }%
  \subfigure[]{%
    \includegraphics[width=0.33\textwidth]{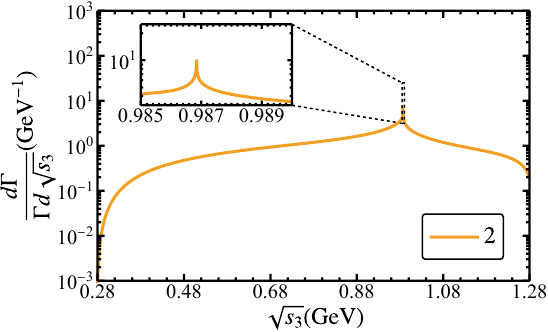}\label{fig:3-2}%
  }%
  \subfigure[]{%
    \includegraphics[width=0.33\textwidth]{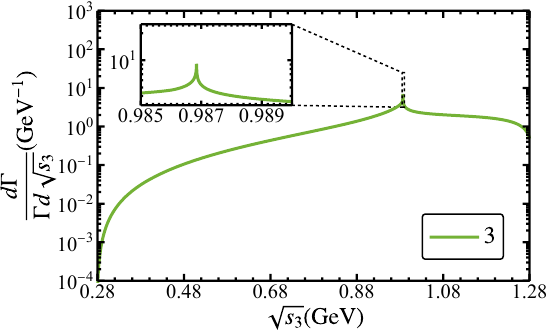}\label{fig:3-3}%
  }
  \subfigure[]{%
    \includegraphics[width=0.33\textwidth]{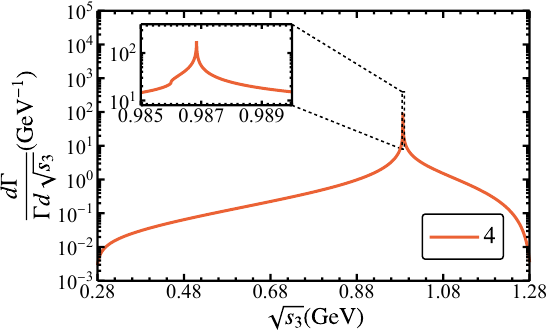}\label{fig:3-4}%
  }%
  \subfigure[]{%
    \includegraphics[width=0.33\textwidth]{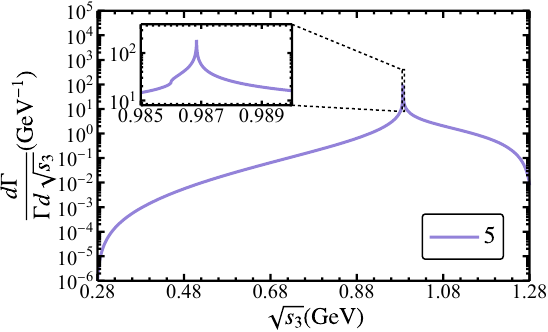}\label{fig:3-5}%
  }%
  \subfigure[]{%
    \includegraphics[width=0.33\textwidth]{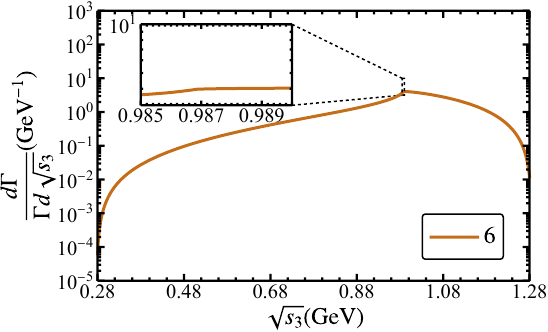}\label{fig:3-6}%
  }
  \subfigure[]{%
    \includegraphics[width=0.33\textwidth]{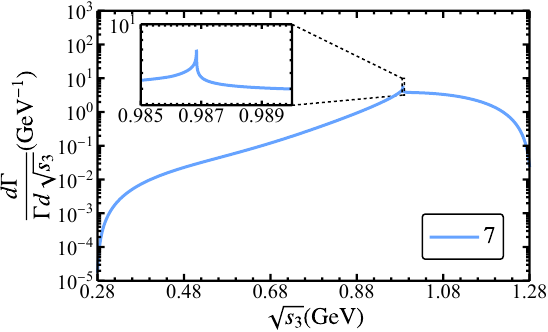}\label{fig:3-7}%
  }%
  \subfigure[]{%
    \includegraphics[width=0.33\textwidth]{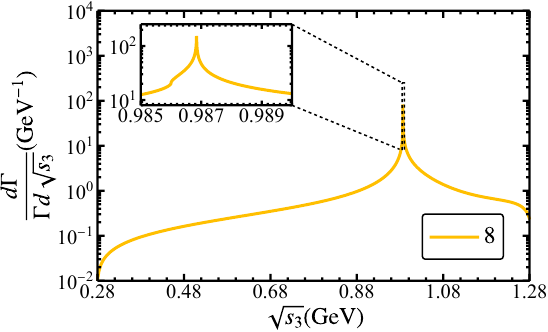}\label{fig:3-8}%
  }%
  \subfigure[]{%
    \includegraphics[width=0.33\textwidth]{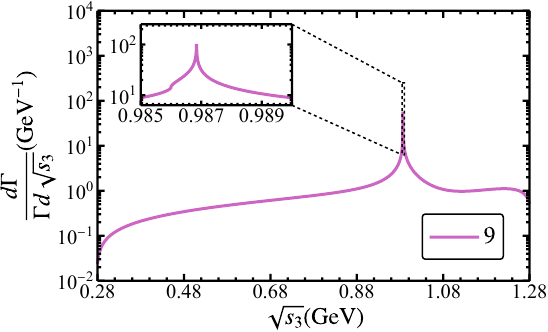}\label{fig:3-9}%
  }
  \subfigure[]{%
    \includegraphics[width=0.33\textwidth]{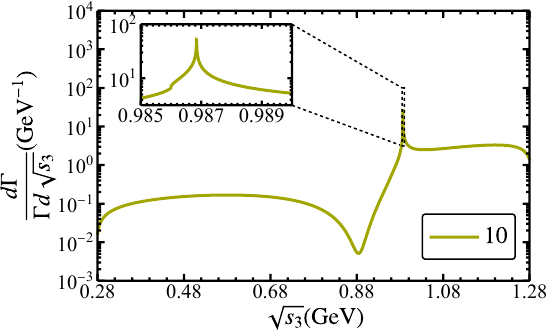}\label{fig:3-10}%
  }%
  \subfigure[]{%
    \includegraphics[width=0.33\textwidth]{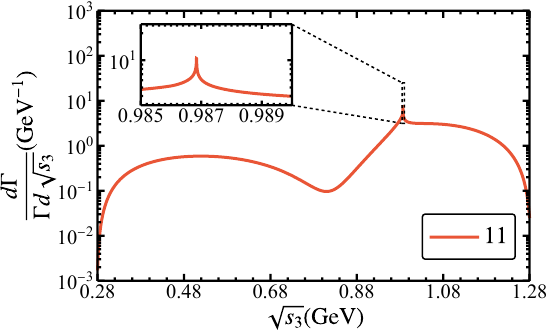}\label{fig:3-11}%
  }%
  \subfigure[]{%
    \includegraphics[width=0.33\textwidth]{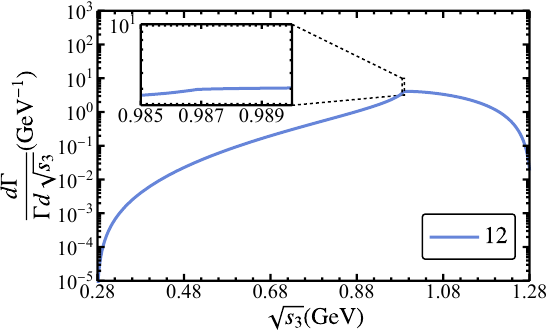}\label{fig:3-12}%
  }%
  \caption{Invariant-mass spectra $\sqrt{s_3}=M(k_1k_2)$ for the same twelve spin-parity assignments and mass inputs as in \cref{fig:spectrum1}, with the final-state resonance factor removed.}   \label{fig:spectrum2}
\end{figure}

To compare the dynamical effects of different vertex structures, we use the masses and kinematics of the $\eta(1405)\to\pi^{+}\pi^{-}\pi^{0}$ system with a $K^{*}\bar{K}K$ loop as a common numerical benchmark for the twelve cases in \cref{tab:possible_quantum_numbers}. Only case $10$ has the spin-parity assignments of that physical decay chain with an intermediate $f_{0}(980)$; the other cases retain the same mass inputs while varying the spins and vertex couplings. The $f_{0}(980)$ mass and width are used as common parameters in the final-state resonance factor, without identifying the intermediate resonance in every case with the physical $f_{0}(980)$. The resulting $k_1k_2$ invariant-mass spectra, with and without this resonance factor, are shown in \cref{fig:spectrum1,fig:spectrum2}.

From these invariant-mass spectra, we see that the prominence of the triangle singularity varies with the choice of vertex couplings. In general, a dominant $C_{0}$ contribution with a nonvanishing Landau-logarithmic coefficient produces a narrow peak. Dynamical effects can suppress this coefficient and weaken the triangle enhancement. The resulting cusp-like profiles must be understood through the remaining ordinary-threshold dependence, smooth background, and interference in the complete amplitude, rather than by assuming that an explicit $B_{0}$ term becomes dominant. We therefore next analyze the scalar-integral contributions together with their coherent threshold expansion to assess the dynamical effects on the triangle singularity.

After the scalar-integral reduction, the coefficients $A_{i}^{j}$, $B_{i}^{j}$, and $C_{i}$ are linear combinations of Lorentz scalars formed from the external momenta and polarization vectors. The coefficients of these Lorentz structures depend on the internal masses and external kinematic invariants:
\begin{equation}
  A_{i}^{j}=\sum_{k}L_{i}^{k}f_{Ai}^{jk},\quad B_{i}^{j}=\sum_{k}L_{i}^{k}f_{Bi}^{jk},\quad C_{i}=\sum_{k}L_{i}^{k}f_{Ci}^{k}\,,
\end{equation}
where $L_{i}^{k}$ is the $k$-th independent Lorentz scalar in the $i$-th diagram, and $f_{Ai}^{jk}$, $f_{Bi}^{jk}$, and $f_{Ci}^{k}$ are its coefficients multiplying the corresponding scalar integrals. By comparing the relative sizes of $\abs{\mathcal{R}[f^{jk}_{Ai}\mathbf{A}_{0j}]}$ and $\abs{\mathcal{R}[f^{jk}_{Bi}\mathbf{B}_{0j}]}$ with $\abs{\mathcal{R}[f^{k}_{Ci}\mathbf{C}_{0}]}$, we can determine the regulated contributions of the different scalar functions and thereby analyze the dynamical effects on the triangle singularity. We first list all the possible Lorentz scalars obtained from the decomposition of the diagrams, as shown in \cref{Lscalar}.
\begin{table}[!ht]
  \centering
  \caption{The possible Lorentz scalars obtained from the decomposition of the diagrams}
  \begin{ruledtabular}    \label{Lscalar}
    \begin{tabular}{cc}
      Serial Number &{$L_{i}^{k}$}\\\hline
      1&$\varepsilon_{1i}\cdot p_{2}$\\
      2&$\varepsilon_{1i}\cdot p_{2},\quad\varepsilon_{1i}\cdot k_1$\\
      3&$\varepsilon_{1i}\cdot p_{2},\quad\varepsilon_{1i}\cdot k_1$\\
      4&$\varepsilon_{\mu\nu\alpha \beta}\varepsilon_{1i}^{\mu}\varepsilon_{1f}^{*\nu}p_{1}^{\alpha}p_{2}^{\beta}$\\
      5&$\varepsilon_{1i}\cdot p_{2}\varepsilon_{1f}^{*}\cdot k_2$\\
      6&$\varepsilon_{\mu\nu\alpha \beta}\varepsilon_{1i}^{\mu}p_2^{\nu}k_1^{\alpha}k_{2}^{\beta}$\\
      7&$\varepsilon_{\mu\nu\alpha \beta}\varepsilon_{1i}^{\mu}p_2^{\nu}k_1^{\alpha}k_{2}^{\beta}$\\
      8&$\varepsilon_{1i}\cdot \varepsilon_{1f}^{*},\quad \varepsilon_{1i}\cdot p_2\varepsilon_{1f}^{*}\cdot p_2,\quad \varepsilon_{1i}\cdot p_2\varepsilon_{1f}^{*}\cdot k_2$\\
      9&$\varepsilon_{1f}^{*}\cdot p_2 ,\quad \varepsilon_{1f}^{*}\cdot k_2$\\
      10&1\\
      11&1\\
      12&1
    \end{tabular}
  \end{ruledtabular}
\end{table}
To compare the contributions of different scalar functions near the triangle singularity, we apply the form-factor subtraction to each complete coefficient-times-integral product and define
\begin{equation}
  a_{i}^{jk}=\frac{\abs{\mathcal{R}[f^{jk}_{Ai}\mathbf{A}_{0j}]}}{\abs{\mathcal{R}[f^{k}_{Ci}\mathbf{C}_{0}]}},\quad b_{i}^{jk}=\frac{\abs{\mathcal{R}[f^{jk}_{Bi}\mathbf{B}_{0j}]}}{\abs{\mathcal{R}[f^{k}_{Ci}\mathbf{C}_{0}]}}\,.
\end{equation}
The numerator and denominator are regulated separately before taking their absolute values. Under the adopted three-propagator mass-subtraction prescription, each $A_{0}$ or $B_{0}$ integral depends on only one or two internal masses, respectively. If the coefficient multiplying such an integral is independent of the internal masses, the complete product is independent of at least one subtracted mass. The corresponding factor $(1-\mathsf{T}_r)$ in \cref{eq:monopole-subtraction} therefore cancels this product exactly. For the amplitudes considered here, all $A_{0}$ coefficients satisfy this condition, giving
\begin{equation}
  \mathcal{R}[f^{jk}_{Ai}\mathbf{A}_{0j}]=0\,.
\end{equation}
This cancellation explains the vanishing $a_i^{jk}$ and some of the exact zeros of $b_i^{jk}$ in \cref{tab:relativeloop}, provided that the common normalization denominator is nonzero. Internal-mass dependence in a coefficient can prevent this cancellation, which is why the subtraction must act on the complete coefficient-times-integral product rather than on the bare scalar integral alone.

\subsection{Threshold and Landau-point expansion}\label{subsec:threshold-expansion}
The normal two-particle threshold associated with the $q_{2}q_{3}$ intermediate state is
\begin{equation}
  s_{\mathrm{th}}=(m_{2}+m_{3})^{2},
\end{equation}
whereas the physical triangle singularity is located at $s_{3}=s_{3}^{-}$. These two points are close in the kinematic region considered here, but they are not identical in general. We therefore introduce a local variable for each expansion point,
\begin{equation}
  \delta=s_{3}-s_{\mathrm{th}},\qquad \Delta=s_{3}-s_{3}^{-},\qquad G(s_{3})=\lambda(s_{1},s_{2},s_{3}).
\end{equation}
The relative momentum of the $q_{2}q_{3}$ pair satisfies
\begin{equation}
  |\mathbf{q}|^{2}=\frac{\lambda(s_{3},m_{2}^{2},m_{3}^{2})}{4s_{3}}=\frac{m_{2}m_{3}}{(m_{2}+m_{3})^{2}}\,\delta+O(\delta^{2}),
\end{equation}
so that this expansion also makes the threshold powers to be associated with the vertex structure explicitly.

At fixed $s_{1}$ and $s_{2}$, the normal threshold and the Landau point require separate local expansions~\cite{Guo:2019twa,Szczepaniak:2015eza}; exact scalar-integral expressions and their analytic continuations are given in Refs.~\cite{tHooft:1978jhc,Denner:1991kt,Ellis:2007qk}. Near an isolated normal threshold, both $\mathbf{B}_{01}$ and $\mathbf{C}_{0}$ carry the $q_{2}q_{3}$ two-particle cut. At threshold, the on-shell momentum is $q_3=m_3p_3/(m_2+m_3)$, so the uncut propagator denominator becomes
\begin{equation}
  d_*\equiv\left.D_1\right|_{q_2q_3\,\mathrm{threshold}}=s_2+m_3^2-m_1^2+\frac{m_3}{m_2+m_3}(s_1-s_2-s_{\mathrm{th}}).
\end{equation}
For $d_*\neq0$, applying the two-particle Cutkosky rule~\cite{Cutkosky:1960sp} shows that the leading triangle discontinuity equals the leading bubble discontinuity divided by $d_*$. Expanding the standard scalar-integral representations~\cite{Denner:1991kt} about this isolated normal threshold, we obtain
\begin{equation}
  \begin{aligned}
    \mathbf{B}_{01}(s_{3})=&\mathbf{B}_{01}^{\mathrm{reg,th}}(s_{3})+\rho_{B}\sqrt{-\delta-i0}+O(|\delta|^{3/2}),\\
    \mathbf{C}_{0}(s_{3})=&\mathbf{C}_{0}^{\mathrm{reg,th}}(s_{3})+\frac{\rho_{B}}{d_*}\sqrt{-\delta-i0}+O(|\delta|^{3/2}).\label{eq:local-B0-C0-expansion}
  \end{aligned}
\end{equation}
The regular parts include the analytic terms at $s_{\mathrm{th}}$, and the normalization-dependent coefficient $\rho_B$ includes the phase appropriate to the chosen square-root branch. The neighborhood of this expansion excludes the distinct Landau point and other simultaneous thresholds. In the following one-variable notation, the dependence on fixed masses, $s_1$, and $s_2$ is suppressed. The remaining independent invariant $s_{23}$ is held fixed unless a smooth kinematic path is specified, and the Lorentz scalars are evaluated in a polarization basis chosen smoothly along the path. For each independent Lorentz scalar, define
\begin{equation}
  \widehat{C}_{i}^{k}(s_{3})=L_{i}^{k}(s_{3})f_{Ci}^{k}(s_{3}),\qquad \widehat{B}_{i}^{k}(s_{3})=L_{i}^{k}(s_{3})f_{Bi}^{1k}(s_{3}).
\end{equation}
Assuming that these coefficients and the external resonance factor are regular at $s_{\mathrm{th}}$, the leading normal-threshold term of the unregulated amplitude is
\begin{equation}
  i\mathcal{M}_{i,k}^{\mathrm{th,nonan}}(s_3)=\frac{g_i\rho_B}{s_{\mathrm{th}}-M_3^2+iM_3\Gamma_3}\left[\widehat{B}_i^k(s_{\mathrm{th}})+\frac{\widehat{C}_i^k(s_{\mathrm{th}})}{d_*}\right]\sqrt{-\delta-i0}+O(|\delta|^{3/2}).\label{eq:normal-threshold-amplitude}
\end{equation}
Thus, its square-root coefficient is a coherent sum, not the coefficient of $\mathbf{B}_{01}$ alone. In particular, with $m_3=m_2$, one has $s_{\mathrm{th}}=4m_2^2$ and $d_*=(s_1+s_2-2m_1^2-2m_2^2)/2$. The coefficients in \cref{tab:coeff6,tab:coeff12} obey
\begin{equation}
  f_{B6}^{11}(s_{\mathrm{th}})+\frac{f_{C6}^{1}(s_{\mathrm{th}})}{d_*}=0,\qquad f_{B12}^{11}(s_{\mathrm{th}})+\frac{f_{C12}^{1}(s_{\mathrm{th}})}{d_*}=0.\label{eq:threshold-cancellation}
\end{equation}
These identities hold for arbitrary $m_1$ and hence also for the subtraction term with $m_1^2\to\Lambda_1^2$. Provided that the regulator-induced singularities lie outside this local region, terms with $m_2$ or $m_3$ shifted have no branch point at the physical $s_{\mathrm{th}}$. The leading square-root term therefore also cancels in the complete subtracted amplitudes $\mathcal{M}_{6,12}$; their normal-threshold nonanalyticity starts at order $|\delta|^{3/2}$ or higher. This cancellation does not remove all threshold dependence.

In a separate neighborhood of a nondegenerate Landau point that excludes $s_{\mathrm{th}}$, the triangle integral has the standard logarithmic singularity~\cite{Landau:1959fi,Coleman:1965xm,Cutkosky:1960sp}, which we write locally as
\begin{equation}
  \mathbf{C}_{0}(s_3)=\mathbf{C}_{0}^{\mathrm{reg,L}}(s_3)+\rho_C\log\left(\frac{\Delta+i0}{\mu^2}\right)+\cdots.\label{eq:local-landau-expansion}
\end{equation}
Here $\rho_C$ is the coefficient of the leading logarithm for the chosen scalar-integral normalization and analytic-continuation convention, and $\mu$ is an arbitrary reference mass that makes the logarithm dimensionless. Changing $\mu$ only shifts the locally regular part. The ordinary-threshold part of $\mathbf{C}_0$ is locally regular in this neighborhood, as are the bubble integrals. Before the form-factor subtraction, the Landau-logarithmic contribution is consequently
\begin{equation}
  i\mathcal{M}_{i,k}^{\mathrm{L,nonan}}(s_3)=\frac{g_i\rho_C\widehat{C}_i^k(s_3)}{s_3-M_3^2+iM_3\Gamma_3}\log\left(\frac{\Delta+i0}{\mu^2}\right)+\cdots.\label{eq:local-nonanalytic-amplitude}
\end{equation}
Equations~\eqref{eq:normal-threshold-amplitude} and \eqref{eq:local-nonanalytic-amplitude} describe different local limits, rather than a single uniform expansion about both branch points. When several Lorentz scalars contribute to the same helicity amplitude, they must be added coherently before taking an absolute value. The ratios $a_{i}^{jk}$ and $b_{i}^{jk}$ used below compare the regulated contributions component by component; they are undefined when $\mathcal{R}[f_{Ci}^{k}\mathbf{C}_{0}]=0$ and require particular care near a zero of this denominator.

\subsection{Relative contributions and lineshape classification}
To evaluate the regulated contributions at a common kinematic point, we specify the remaining invariant of the $1\to 3$ process. At fixed initial and final particle masses, there are two independent Dalitz invariants, which we choose as $s_{3}=(k_1+k_2)^{2}$ and $s_{23}=(p_2+k_2)^{2}$. At $s_{3}=s_{3}^{-}$, the allowed range of $s_{23}$ is
\begin{equation}
  \begin{aligned}
    s_{23}^{\text{min}}=&(\tilde{E}_{k_2}+\tilde{E}_{p_2})^{2}-\qty(\sqrt{\tilde{E}^{2}_{k_2}-k_2^{2}}+\sqrt{\tilde{E}^{2}_{p_2}-s_2})^{2} \\
    s_{23}^{\text{max}}=&(\tilde{E}_{k_2}+\tilde{E}_{p_2})^{2}-\qty(\sqrt{\tilde{E}^{2}_{k_2}-k_2^{2}}-\sqrt{\tilde{E}^{2}_{p_2}-s_2})^{2}\,,
  \end{aligned}
\end{equation}
where
\begin{equation}
  \tilde{E}_{k_2}=\frac{1}{2\sqrt{s_3}}(s_3+k_2^{2}-k_1^{2}),\quad \tilde{E}_{p_2}=\frac{1}{2\sqrt{s_3}}(p_1^{2}-s_3-s_2)\,.
\end{equation}
To numerically compute $a_{i}^{jk}$ and $b_{i}^{jk}$, we take
\begin{equation}
  s_{23}=\frac{s_{23}^{\text{min}}+s_{23}^{\text{max}}}{2}=2\tilde{E}_{k_2}\tilde{E}_{p_2}+k_2^{2}+s_2\,.\label{eq:s23ts}
\end{equation}
The results for $a_{i}^{jk}$, $b_{i}^{jk}$, and $\abs*{\mathcal{R}[f^{k}_{Ci}\mathbf{C}_{0}]}$ are shown in \cref{tab:relativeloop}.
\begin{table}[!ht]
  \centering
  \caption{Numerical results for $a_{i}^{jk}$, $b_{i}^{jk}$, and $\abs{\mathcal{R}[f^{k}_{Ci}\mathbf{C}_{0}]}$ with the form factor included. Parentheses list the independent Lorentz components in the order of \cref{Lscalar}.}
  \resizebox{\textwidth}{!}{
    \begin{ruledtabular}    \label{tab:relativeloop}
      \begin{tabular}{cccccccc}
        Serial  &$a_{i}^{1k}$&$a_{i}^{2k}$&$a_{i}^{3k}$& $b_{i}^{1k}$& $b_{i}^{2k}$& $b_{i}^{3k}$&$\abs{\mathcal{R}[f^{k}_{Ci}\mathbf{C}_{0}]}$\\\hline
        1&0&0&0&0&0&0&$2.05$\\[1ex]
        \multirow{2}{*}{2}&\multirow{2}{*}{$\mqty(0\\0)$}&\multirow{2}{*}{$\mqty(0\\0)$}&\multirow{2}{*}{$\mqty(0\\0)$}&\multirow{2}{*}{$\mqty(0.438\\0.438)$}&\multirow{2}{*}{$\mqty(4.85\times 10^{-4}\\4.85\times 10^{-4})$}&\multirow{2}{*}{$\mqty(0.483\\0.483)$}&\multirow{2}{*}{$\mqty(2.37\times 10^{-2}\\4.74\times 10^{-2})$}\\\\[1ex]
        \multirow{2}{*}{3}&\multirow{2}{*}{$\mqty(0\\0)$}&\multirow{2}{*}{$\mqty(0\\0)$}&\multirow{2}{*}{$\mqty(0\\0)$}&\multirow{2}{*}{$\mqty(0.301\\0.301)$}&\multirow{2}{*}{$\mqty(6.36\times 10^{-3}\\6.36\times 10^{-3})$}&\multirow{2}{*}{$\mqty(0.332\\0.332)$}&\multirow{2}{*}{$\mqty(2.56\times 10^{-2}\\5.12\times 10^{-2})$}\\\\[1ex]
        4&0&0&0&0&0&0&$2.05$\\
        5&0&0&0&$4.43\times 10^{-3}$&$9.36\times 10^{-5}$&$6.40\times 10^{-3}$&$14.3$\\
        6&0&0&0&$0.438$&$4.85\times 10^{-4}$&$0.483$&$9.48\times 10^{-2}$\\
        7&0&0&0&0&$2.98\times 10^{-2}$&$0.268$&$4.50\times 10^{-2}$\\
        \multirow{3}{*}{8}&\multirow{3}{*}{$\mqty(0\\0\\0)$}&\multirow{3}{*}{$\mqty(0\\0\\0)$}&\multirow{3}{*}{$\mqty(0\\0\\0)$}&\multirow{3}{*}{$\mqty(1.80\times 10^{-2}\\3.39\times 10^{-2}\\1.19\times 10^{-2})$}&\multirow{3}{*}{$\mqty(1.99\times 10^{-5}\\2.63\times 10^{-4}\\1.00\times 10^{-5})$}&\multirow{3}{*}{$\mqty(1.98\times 10^{-2}\\3.74\times 10^{-2}\\9.98\times 10^{-3})$}&\multirow{3}{*}{$\mqty(0.288\\2.79\\3.73)$}\\\\\\[1ex]
        \multirow{2}{*}{9}&\multirow{2}{*}{$\mqty(0\\0)$}&\multirow{2}{*}{$\mqty(0\\0)$}&\multirow{2}{*}{$\mqty(0\\0)$}&\multirow{2}{*}{$\mqty(0\\0)$}&\multirow{2}{*}{$\mqty(0\\0)$}&\multirow{2}{*}{$\mqty(0\\0)$}&\multirow{2}{*}{$\mqty(1.37\\0.684)$}\\\\[1ex]
        10&0&0&0&0&0&0&$0.291$\\[1ex]
        11&0&0&0&$5.94\times 10^{-2}$&$0.195$&$4.11\times 10^{-2}$&$4.97\times 10^{-16}$\\[1ex]
        12&0&0&0&$1.34\times 10^{-3}$&$8.13\times 10^{-4}$&$4.66\times 10^{-2}$&$2.95\times 10^{-16}$
      \end{tabular}
  \end{ruledtabular}}
\end{table}

The very small absolute $C_0$ contributions in \cref{tab:relativeloop} for the cases illustrated in \cref{fig:2-11,fig:2-12} result from the midpoint choice of $s_{23}$. For $k_1^2=k_2^2=M_f^2$, \cref{eq:s23ts} gives
\begin{equation}
  s_{23}^{\mathrm{mid}}=M_f^2+\frac{s_1+s_2-s_3}{2},\qquad H\equiv2M_f^2+s_1+s_2-s_3-2s_{23}=0\quad\text{at }s_{23}=s_{23}^{\mathrm{mid}}.\label{eq:dalitz-midpoint-zero}
\end{equation}
The coefficients of $\mathcal{M}_{11,12}$ contain this common factor $H$, which is independent of the propagator masses and hence is preserved by the form-factor subtraction. The values $4.97\times10^{-16}$ and $2.95\times10^{-16}$ are numerical residuals of this kinematic zero, rather than additional dynamical suppression scales. This point-specific zero does not imply vanishing amplitudes away from the midpoint or vanishing invariant-mass spectra. The qualitative lineshape classification is based on the full amplitudes and spectra, not on these residual magnitudes, and is therefore unchanged by this special choice of $s_{23}$.

The separate local expansions in \cref{eq:normal-threshold-amplitude,eq:local-nonanalytic-amplitude} identify the coherent normal-threshold term and the Landau logarithm, respectively, whereas \cref{tab:relativeloop} compares the complete coefficient-times-integral products after the form-factor subtraction. The tabulated $b_i^{1k}$ uses $\mathcal{R}[f_{Bi}^{1k}\mathbf{B}_{01}]$ and $\mathcal{R}[f_{Ci}^{k}\mathbf{C}_{0}]$, including their regular parts, and is not a ratio of the leading nonanalytic terms. A zero of the physical-mass coefficient $f_{Ci}^{k}$ does not require the complete subtracted $C_{0}$ contribution to vanish: the shifted-mass terms can retain a regular contribution at the physical Landau point. Likewise, a vanishing explicit $B_{01}$ contribution after subtraction does not remove the ordinary two-particle threshold, which can also be carried by the $C_{0}$ combination. At fixed $s_1$ and $s_2$, the $B_{02}$ and $B_{03}$ terms contribute to the local smooth background, while the $A_{0}$ terms cancel exactly. The table therefore shows the relative sizes of the scalar-integral contributions at the selected kinematic point, subject to the midpoint qualification above, while their energy dependence and interference determine the full lineshape.

The three classes of lineshapes can be understood in light of \cref{tab:relativeloop} as follows:
\begin{enumerate}[(I)]
  \item For serial numbers $1,4,5,8,9$, and $10$, the regulated $C_{0}$ term dominates over the explicit $B_{0}$ terms. All three $b_i^{jk}$ vanish for $i=1,4,9,10$, while the largest bubble ratios for the cases shown in \cref{fig:2-5,fig:2-8} are only $6.40\times10^{-3}$ and $3.74\times10^{-2}$, respectively. The small bubble contributions allow the logarithmic enhancement associated with the triangle integral to remain prominent. This is consistent with the pronounced narrow TS peaks shown in \cref{fig:2-1,fig:2-4,fig:2-5,fig:2-8,fig:2-9,fig:2-10} and \cref{fig:3-1,fig:3-4,fig:3-5,fig:3-8,fig:3-9,fig:3-10}.
  \item For serial numbers $2,3,7$, and $11$, the spectra exhibit less pronounced TS peaks, and the bubble contributions can compete more strongly with the triangle term than in class (I). For both Lorentz components of the case shown in \cref{fig:2-2}, the $B_{01}$ and $B_{03}$ ratios are approximately $0.438$ and $0.483$, respectively; the corresponding ratios for the case shown in \cref{fig:2-3} are $0.301$ and $0.332$. For the case shown in \cref{fig:2-7}, $b_7^{11}=0$, but $b_7^{31}\simeq0.268$ gives a sizable smooth contribution. These entries indicate stronger competition between the scalar-integral terms than in class (I). The tabulated ratios for the case shown in \cref{fig:2-11} are $b_{11}^{11}\simeq5.94\times10^{-2}$, $b_{11}^{21}\simeq0.195$, and $b_{11}^{31}\simeq4.11\times10^{-2}$; because of the midpoint zero discussed above, these ratios of numerical residuals are not used on their own to establish its lineshape classification. The full $B_{01}$ and $C_0$ terms share the ordinary-threshold cut, while the smooth $B_{02}$ and $B_{03}$ contributions can interfere with the triangle term and reduce the prominence of the TS peak, without requiring the Landau logarithm to disappear from every Lorentz component. This interpretation is consistent with the less pronounced peaks in \cref{fig:2-2,fig:2-3,fig:2-7,fig:2-11} and \cref{fig:3-2,fig:3-3,fig:3-7,fig:3-11}. For generic $s_{23}$, the coefficient $f_{C2}^{2}$ contains a Landau-zero factor. At the selected midpoint, $f_{C2}^{1}$ also vanishes at the physical Landau point, so both physical-mass $C_{0}$ coefficients are zero there. These zeros must be distinguished from the finite subtracted $C_{0}$ contributions in the table and do not determine the integrated lineshape by themselves.
  \item For serial numbers $6$ and $12$, the distinguishing dynamical feature is the Landau-zero factor in the physical-mass $C_{0}$ coefficient, which suppresses the leading logarithmic enhancement. For the case shown in \cref{fig:2-6}, the full subtracted $C_{0}$ contribution remains finite, with magnitude $9.48\times10^{-2}$, while $b_6^{11}\simeq0.438$ and $b_6^{31}\simeq0.483$. For the case shown in \cref{fig:2-12}, the tabulated $C_{0}$ contribution is $2.95\times10^{-16}$, and the quoted bubble ratios are $1.34\times10^{-3}$, $8.13\times10^{-4}$, and $4.66\times10^{-2}$. Its tiny absolute contribution reflects the special $s_{23}$ choice described above and must be distinguished from the dynamical Landau zero. Neither the ratios for the case shown in \cref{fig:2-6} nor the residual ratios for the case shown in \cref{fig:2-12} establish dominance of an explicit $B_0$ term. Moreover, \cref{eq:threshold-cancellation} shows that the leading ordinary-threshold square roots cancel between the $B_{01}$ and $C_0$ contributions in these amplitudes. The cusp-like profiles in \cref{fig:2-6,fig:2-12} and \cref{fig:3-6,fig:3-12} should therefore be understood through the weakened Landau enhancement, the higher-order ordinary-threshold dependence, and interference with the smooth background of the complete amplitude, rather than a large $B_0/C_0$ ratio or an isolated $B_{01}$ square-root term.
\end{enumerate}

\subsection{Coefficient zeros and regulated loop contributions}

\begin{figure}[!ht]
    \centering
    \subfigure[]{%
      \includegraphics[width=0.33\textwidth]{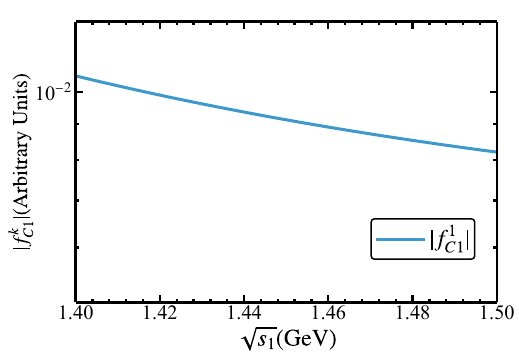}\label{fig:4-1}%
    }%
    \subfigure[]{%
      \includegraphics[width=0.33\textwidth]{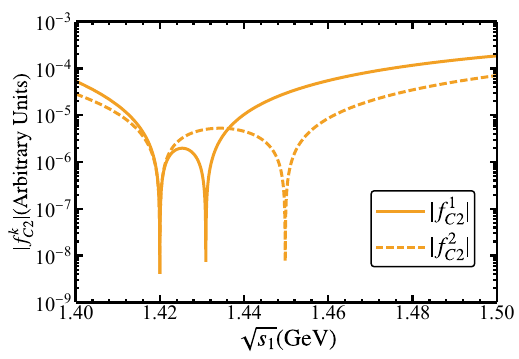}\label{fig:4-2}%
    }%
    \subfigure[]{%
      \includegraphics[width=0.33\textwidth]{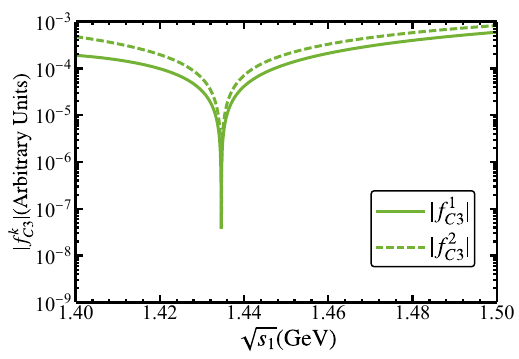}\label{fig:4-3}%
    }
    \subfigure[]{%
      \includegraphics[width=0.33\textwidth]{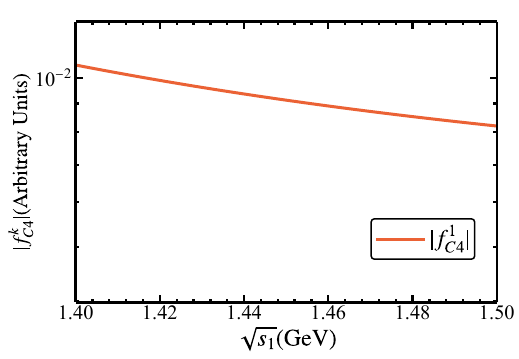}\label{fig:4-4}%
    }%
    \subfigure[]{%
      \includegraphics[width=0.33\textwidth]{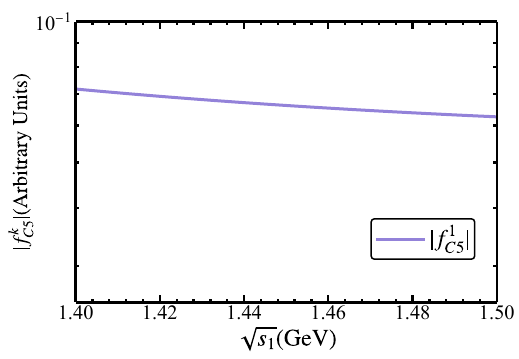}\label{fig:4-5}%
    }%
    \subfigure[]{%
      \includegraphics[width=0.33\textwidth]{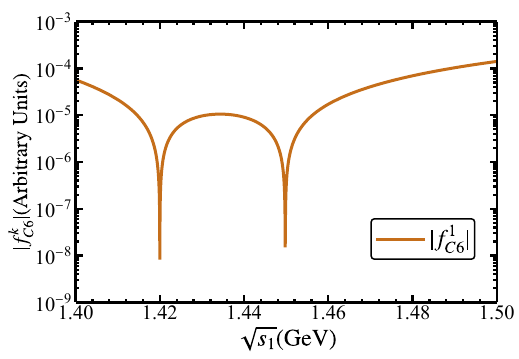}\label{fig:4-6}%
    }
    \subfigure[]{%
      \includegraphics[width=0.33\textwidth]{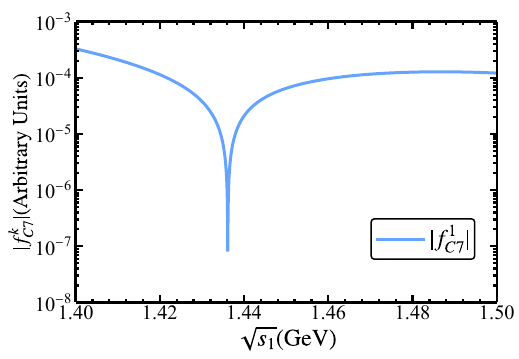}\label{fig:4-7}%
    }%
    \subfigure[]{%
      \includegraphics[width=0.33\textwidth]{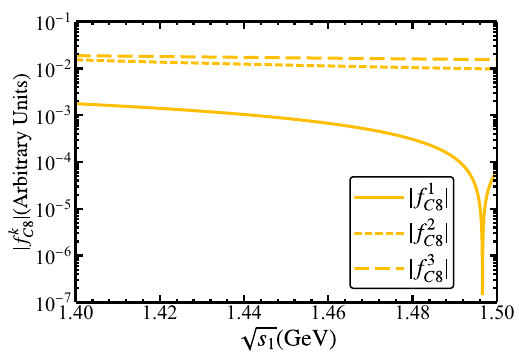}\label{fig:4-8}%
    }%
    \subfigure[]{%
      \includegraphics[width=0.33\textwidth]{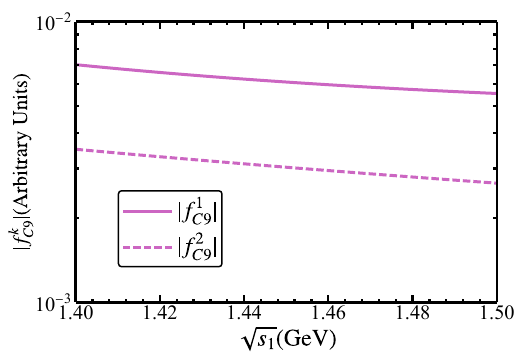}\label{fig:4-9}%
    }
    \subfigure[]{%
      \includegraphics[width=0.33\textwidth]{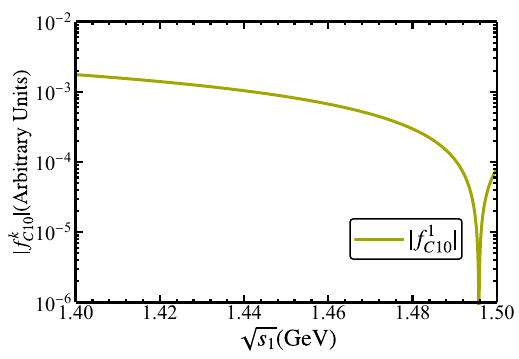}\label{fig:4-10}%
    }%
    \subfigure[]{%
      \includegraphics[width=0.33\textwidth]{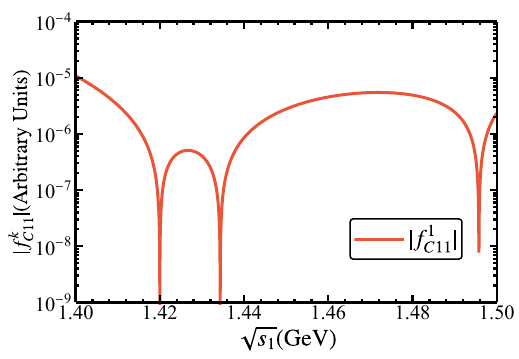}\label{fig:4-11}%
    }%
    \subfigure[]{%
      \includegraphics[width=0.33\textwidth]{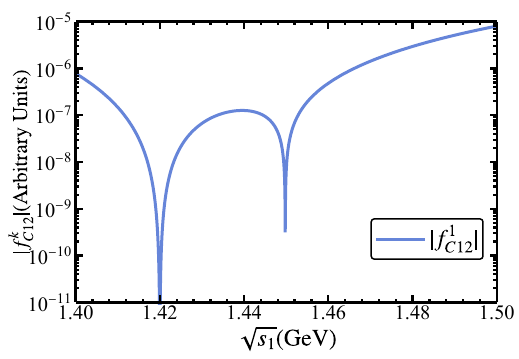}\label{fig:4-12}%
    }%
    \caption{Magnitudes of the physical-mass $C_{0}$ coefficients $f_{Ci}^{k}$ before the form-factor subtraction, as functions of the initial-state center-of-mass energy.}   \label{fig:C0coeff}
  \end{figure}

\begin{figure}[!ht]
    \centering
    \subfigure[]{%
      \includegraphics[width=0.33\textwidth]{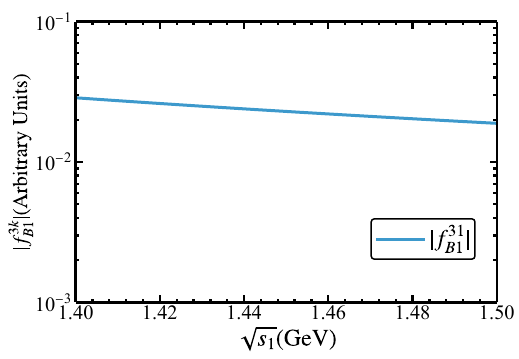}\label{fig:5-1}%
    }%
    \subfigure[]{%
      \includegraphics[width=0.33\textwidth]{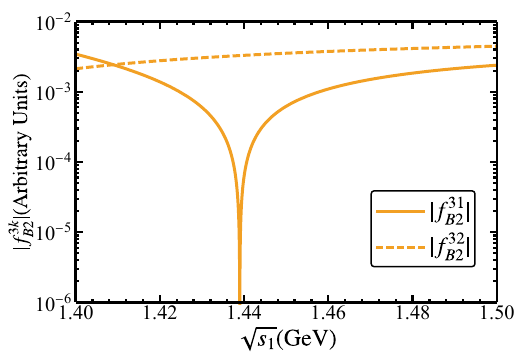}\label{fig:5-2}%
    }%
    \subfigure[]{%
      \includegraphics[width=0.33\textwidth]{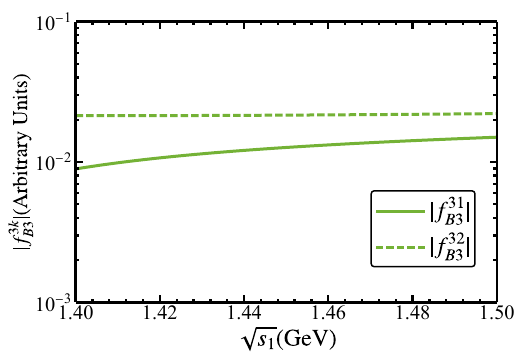}\label{fig:5-3}%
    }
    \subfigure[]{%
      \includegraphics[width=0.33\textwidth]{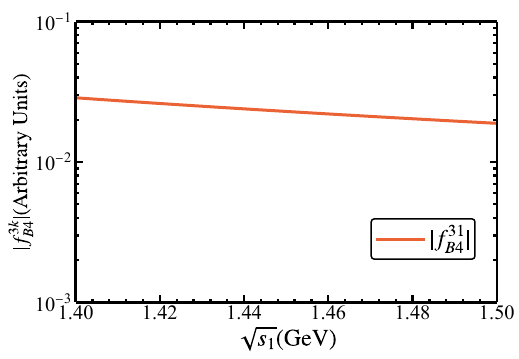}\label{fig:5-4}%
    }%
    \subfigure[]{%
      \includegraphics[width=0.33\textwidth]{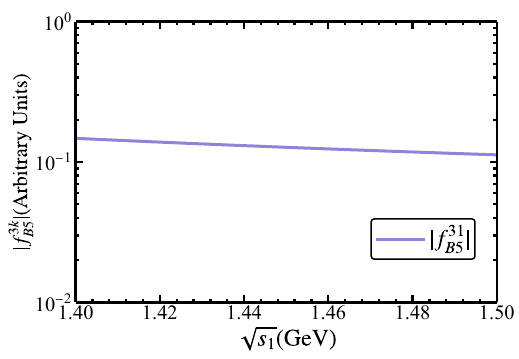}\label{fig:5-5}%
    }%
    \subfigure[]{%
      \includegraphics[width=0.33\textwidth]{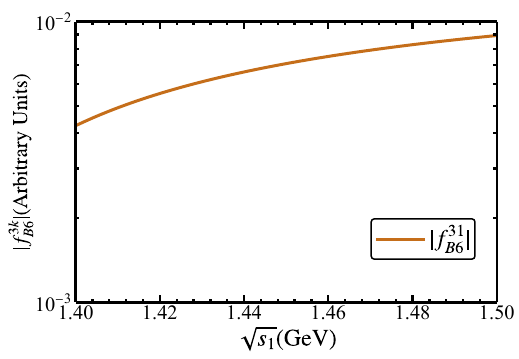}\label{fig:5-6}%
    }
    \subfigure[]{%
      \includegraphics[width=0.33\textwidth]{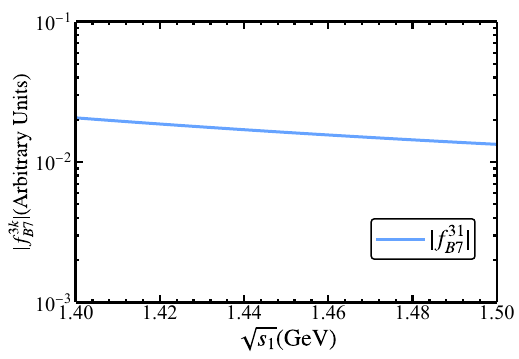}\label{fig:5-7}%
    }%
    \subfigure[]{%
      \includegraphics[width=0.33\textwidth]{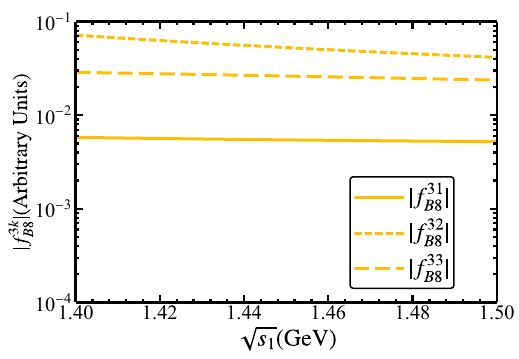}\label{fig:5-8}%
    }%
    \subfigure[]{%
      \includegraphics[width=0.33\textwidth]{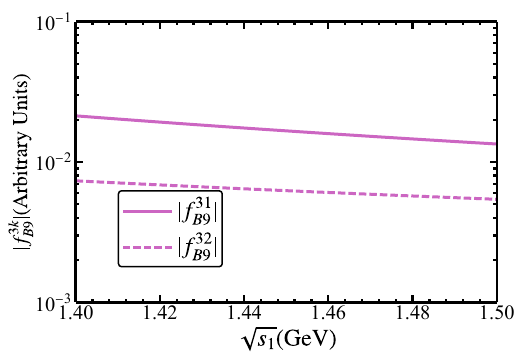}\label{fig:5-9}%
    }
    \subfigure[]{%
      \includegraphics[width=0.33\textwidth]{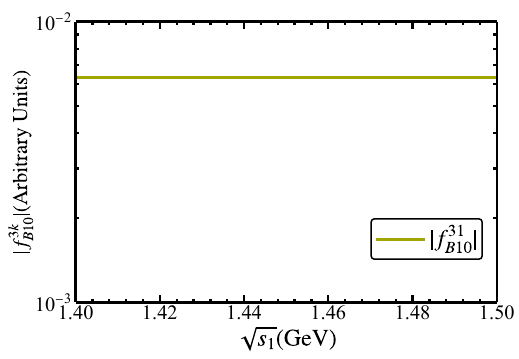}\label{fig:5-10}%
    }%
    \subfigure[]{%
      \includegraphics[width=0.33\textwidth]{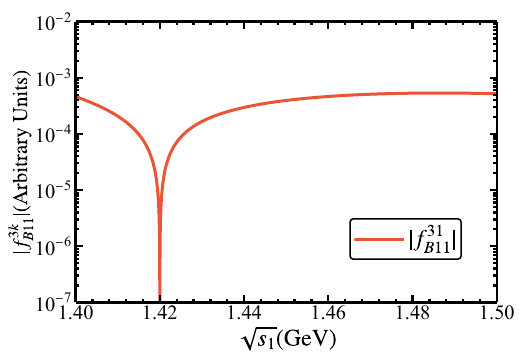}\label{fig:5-11}%
    }%
    \subfigure[]{%
      \includegraphics[width=0.33\textwidth]{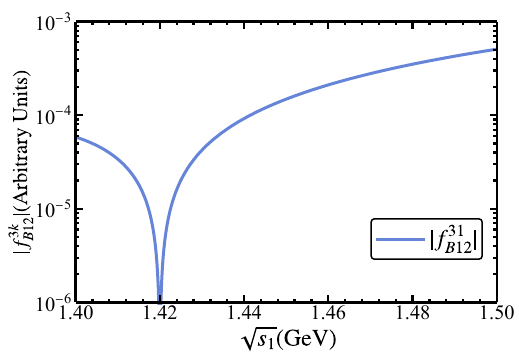}\label{fig:5-12}%
    }%
    \caption{Magnitudes of the physical-mass coefficients $f_{Bi}^{3k}$ of $\mathbf{B}_{03}$ before the form-factor subtraction, as functions of the initial-state center-of-mass energy.}   \label{fig:B0coeff}
  \end{figure}

\begin{figure}[!ht]
  \centering
  \subfigure[]{%
    \includegraphics[width=0.43\textwidth]{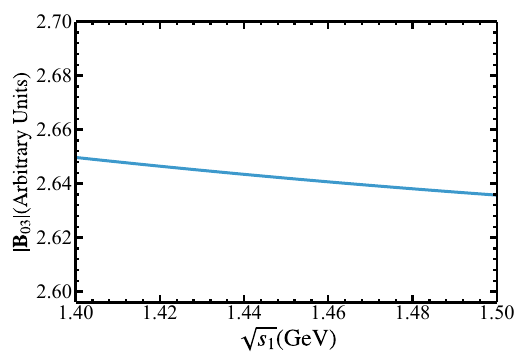}\label{fig:6-1}%
  }%
  \subfigure[]{%
    \includegraphics[width=0.43\textwidth]{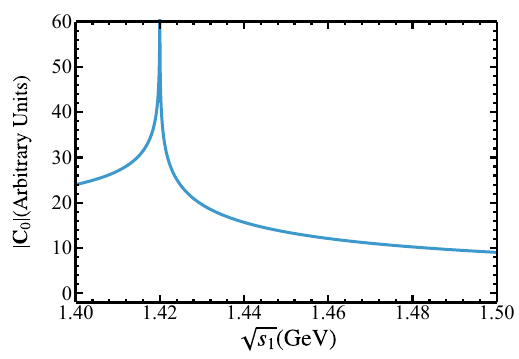}\label{fig:6-2}%
  }%
  \caption{Evolution of $\abs{\mathbf{B}_{03}}$ and $\abs{\mathbf{C}_{0}}$ with the initial-state center-of-mass energy, evaluated at the physical masses without the form-factor subtraction.}
  \label{fig:B0C0}
\end{figure}

  We examine the zeros of all seventeen physical-mass $C_0$ coefficients before the form-factor subtraction, distinguishing Landau factors from kinematic and numerator cancellations. Their initial-energy scans in \cref{fig:C0coeff} are then compared with the $B_{03}$ coefficients and scalar integrals in \cref{fig:B0coeff,fig:B0C0}.

  We use the component ordering in \cref{Lscalar} and the assumptions $m_3=m_2$ and $k_1^2=k_2^2=M_f^2$ of Appendix~\ref{appendix:A}. The common variables are
  \begin{equation}
    d=m_1^2-m_2^2,\qquad U=s_1+s_2,\qquad V=s_1-s_2,\qquad Z=s_3,\qquad G=V^2-2UZ+Z^2.\label{eq:remaining-coefficient-variables}
  \end{equation}
  The factor $H$ is defined in \cref{eq:dalitz-midpoint-zero}. Here $s_1$, $s_2$, $s_3$, and $s_{23}$ remain independent invariants; no midpoint condition is imposed. The Landau polynomial entering the coefficient decompositions is
  \begin{equation}
    \mathcal{P}(s_{1},s_{2},s_{3})=m_{2}^{4}s_{3}+\left[m_{1}^{4}+s_{1}s_{2}-m_{1}^{2}(s_{1}+s_{2}-s_{3})\right]s_{3}+m_{2}^{2}\left[(s_{1}-s_{2})^{2}-(2m_{1}^{2}+s_{1}+s_{2})s_{3}\right].\label{eq:landau-factor}
  \end{equation}
  For $m_3=m_2$, $\mathcal{P}$ equals minus four times the polynomial on the right-hand side of \cref{eq:laeq}. Its two roots in $s_3$ are
  \begin{equation}
    s_{3}^{\pm}=4m_{2}^{2}+\frac{(m_{1}^{2}+m_{2}^{2}-s_{2})(s_{1}-m_{1}^{2}-m_{2}^{2})-4m_{1}^{2}m_{2}^{2}\pm\sqrt{\lambda(s_{1},m_{1}^{2},m_{2}^{2})\lambda(s_{2},m_{1}^{2},m_{2}^{2})}}{2m_{1}^{2}},\label{eq:landau-roots}
  \end{equation}
  where $s_3^-$ is the branch satisfying the Coleman--Norton condition in the kinematic window considered here. Thus,
  \begin{equation}
    \mathcal{P}(s_{1},s_{2},s_{3})=m_{1}^{2}(s_{3}-s_{3}^{-})(s_{3}-s_{3}^{+}).\label{eq:landau-factorization}
  \end{equation}
  To display the other zero conditions compactly, we also define two linear factors,
  \begin{equation}
    \Xi=U-Z-2d,\qquad \Omega=U-2Z-d+m_2^2,\label{eq:remaining-linear-factors}
  \end{equation}
  and five auxiliary polynomials,
  \begin{align}
    \mathcal{Q}&=Z(s_1+2s_2-2Z-3d)+2m_2^2(V+Z),\label{eq:remaining-Q}\\
    \mathcal{J}&=-d(V-Z)+s_2(V+Z),\label{eq:remaining-J}\\
    \mathcal{N}&=G+(d-s_2)(V+Z)-Z(U-Z),\label{eq:remaining-N}\\
    \mathcal{K}&=G\mathcal{Q}-(V+Z)\mathcal{P},\label{eq:remaining-K}\\
    \mathcal{W}&=m_2^2(Z^2-V^2)+\frac{Z}{2}\left[-6dV+4dZ+3UV+9UZ-5V^2-3VZ-4Z^2\right].\label{eq:remaining-W}
  \end{align}

  With these definitions, the coefficients sharing $\mathcal{N}$ take the forms
  \begin{align}
    f_{C1}^{1}&=-\frac{\mathcal{N}}{G},\qquad f_{C4}^{1}=\frac{\mathcal{N}}{G},\label{eq:fC14-complete}\\
    f_{C5}^{1}&=\frac{2}{G}\left[(d+3V-3Z)\mathcal{N}+2G(-d+2m_2^2-s_1+2s_2+2Z)\right].\label{eq:fC5-complete}
  \end{align}
  The two components of each of $f_{C2}$ and $f_{C3}$ are
  \begin{align}
    f_{C2}^{1}&=\frac{[G+2H(V+Z)]\mathcal{P}-H\mathcal{K}}{G^2},\label{eq:fC21-complete}\\
    f_{C2}^{2}&=\frac{2\mathcal{P}}{G},\label{eq:fC2-factor}\\
    f_{C3}^{1}&=\frac{[G+H(V+Z)]\mathcal{K}-8Hs_1Z\mathcal{P}}{4G^2},\label{eq:fC31-complete}\\
    f_{C3}^{2}&=\frac{\mathcal{K}}{2G}.\label{eq:fC32-complete}
  \end{align}
  The next two coefficients distinguish an overall Landau factor from a difference of terms:
  \begin{align}
    f_{C6}^{1}&=-\frac{4\mathcal{P}}{G},\label{eq:fC6-factor}\\
    f_{C7}^{1}&=\frac{2(Z\Xi\Omega-2\mathcal{P})}{G}.\label{eq:fC7-complete}
  \end{align}
  The three components of $f_{C8}$ are
  \begin{align}
    f_{C8}^{1}&=\frac{\mathcal{P}-2G\Omega}{2G},\label{eq:fC81-complete}\\
    f_{C8}^{2}&=\frac{3(V+Z)\mathcal{P}-G(\mathcal{Q}+4\mathcal{J})}{G^2},\label{eq:fC82-complete}\\
    f_{C8}^{3}&=-\frac{[G+3Z(U-Z)]\mathcal{P}+G\mathcal{W}}{ZG^2}.\label{eq:fC83-complete}
  \end{align}
  Finally, the two components of $f_{C9}$ and the coefficients $f_{C10}^{1}$, $f_{C11}^{1}$, and $f_{C12}^{1}$ read
  \begin{align}
    f_{C9}^{1}&=\frac{G-Z\Xi}{G},\qquad f_{C9}^{2}=\frac{\mathcal{J}}{G},\label{eq:fC9-complete}\\
    f_{C10}^{1}&=\Omega,\qquad f_{C11}^{1}=\frac{ZH\Xi\Omega}{G},\label{eq:fC1011-complete}\\
    f_{C12}^{1}&=-\frac{ZH\mathcal{P}}{G}.\label{eq:fC12-factor}
  \end{align}

  On the $q_2q_3$ cut in the rest frame of $p_3$, let $z$ be the cosine of the angle between $\boldsymbol p_2$ and $\boldsymbol q_3$. The uncut denominator is
  \begin{equation}
    D_1=a-bz,\qquad a=\frac{\Xi}{2},\qquad b=2|\boldsymbol p_2||\boldsymbol q_3|=\frac{\sqrt{G\lambda(Z,m_2^2,m_2^2)}}{2Z}.\label{eq:cut-angular-denominator}
  \end{equation}
  For $b\neq0$, its angular pole lies at $z_0=a/b$, and
  \begin{equation}
    \mathcal{P}=Z(a^2-b^2)=Zb^2(z_0^2-1).\label{eq:landau-angular-endpoints}
  \end{equation}
  Thus, $\mathcal{P}=0$ places the pole at a collinear endpoint $z_0=\pm1$. A physical TS also requires positive Feynman parameters and the Coleman--Norton configuration; it is a logarithmic branch point, not a resonance pole.

  The overall factor $\mathcal{P}$ in $f_{C2}^{2}$, $f_{C6}^{1}$, and $f_{C12}^{1}$ suppresses their leading Landau logarithms at generic $s_{23}$. The first component of $f_{C2}$ shares this zero at the midpoint:
  \begin{equation}
    \left.f_{C2}^{1}\right|_{H=0}=\frac{\mathcal{P}}{G}=\frac{1}{2}f_{C2}^{2}.\label{eq:fC2-midpoint}
  \end{equation}
  In contrast, $f_{C11}^{1}$ contains the external midpoint factor $H$ and the mass-dependent factors $\Xi$ and $\Omega$, the latter shared with $f_{C10}^{1}$.

  The common vertex contraction in the amplitudes associated with \cref{fig:4-10,fig:4-11} satisfies
  \begin{equation}
    \left.(p_2-q_3)\cdot(p_1+q_2)\right|_{D_1=D_2=D_3=0}=s_1+s_2-2Z-m_1^2+2m_2^2=\Omega.\label{eq:omega-onshell-contraction}
  \end{equation}
  Hence $\Omega=0$ is a numerator-contraction zero. For $f_{C11}^{1}$, the azimuthal average of $(q_3-q_2)\cdot(k_1-k_2)$ on the two-particle cut is $H|\boldsymbol q_3|z/|\boldsymbol p_2|$, which becomes $ZH\Xi/G$ at $z=z_0$. Thus, $\Xi=0$ is a longitudinal-projection zero at the angular midpoint $z_0=0$, not a new endpoint singularity. Neither factor by itself establishes a physical TS.

  Other zeros follow from numerator cancellations. The pair $f_{C1}^{1},f_{C4}^{1}$ shares $\mathcal{N}=0$, whereas $f_{C5}^{1}$ requires cancellation between the two terms in \cref{eq:fC5-complete}. The $H$-dependent terms likewise allow $f_{C2}^{1}$ and $f_{C3}^{1}$ to vanish away from the roots of $\mathcal{P}$ and $\mathcal{K}$, respectively. Although $f_{C3}^{1}=f_{C3}^{2}/2$ at $H=0$, their zeros generally differ because
  \begin{equation}
    \left.f_{C3}^{1}\right|_{\mathcal{K}=0}=-\frac{2Hs_1Z\mathcal{P}}{G^2},\label{eq:fC3-separated-zeros}
  \end{equation}
  which need not vanish. The distinct cancellation conditions for $f_{C7}^{1}$, $f_{C8}^{1,2,3}$, and $f_{C9}^{1,2}$ follow directly from \cref{eq:fC7-complete,eq:fC81-complete,eq:fC82-complete,eq:fC83-complete,eq:fC9-complete}. These expressions assume $G\neq0$; at Gram degeneracy the complete helicity amplitude must be assembled before taking the limit. The apparent denominator factor $Z$ in $f_{C8}^{3}$ is removable.

  At fixed $s_1,s_2$, the overall Landau factor behaves near $\Delta=s_3-s_3^-=0$ as
  \begin{equation}
    \mathcal{P}=m_{1}^{2}(s_{3}^{-}-s_{3}^{+})\Delta+O(\Delta^{2}).\label{eq:landau-local}
  \end{equation}
  With smooth nonzero prefactors, it changes $\log[(\Delta+i0)/\mu^2]$ into $\Delta\log[(\Delta+i0)/\mu^2]$. An additional $H$ zero can increase the suppression along a special path, but neither cancellation creates a new pole nor necessarily removes all nonanalytic dependence.

  The scan in \cref{fig:C0coeff} varies $s_1$ at fixed $s_2,s_3,s_{23}$, using $m_1=0.891~\mathrm{GeV}$, $m_2=m_3=0.493~\mathrm{GeV}$, $\sqrt{s_2}=0.135~\mathrm{GeV}$, and $M_f=0.139~\mathrm{GeV}$. At $s_{1*}=(1.42~\mathrm{GeV})^2$, \cref{eq:s3ts} gives $s_{3*}\simeq0.973861381~\mathrm{GeV}^2$ and the midpoint is $s_{23*}\simeq0.549702810~\mathrm{GeV}^2$. Holding these values fixed yields
  \begin{equation}
    s_{23}=s_{23*}=M_f^2+\frac{s_{1*}+s_2-s_{3*}}{2},\qquad H=s_1-s_{1*}.\label{eq:fixed-s23-scan}
  \end{equation}
  Thus, the scan does not follow the moving midpoint: changing $s_{23*}$ moves the $H$ zero but not the Landau singularity. Roots are evaluated using the exact Landau definition of $s_{3*}$; $G,Z>0$ throughout the plotted interval. This $s_1$ scan is distinct from the $s_3$ spectra in \cref{fig:spectrum1,fig:spectrum2}.

  The normal $q_1q_2$ threshold shared by $\mathbf{B}_{03}$ and $\mathbf{C}_0$ is $\sqrt{s_{1,\mathrm{th}}}=m_1+m_2=1.384~\mathrm{GeV}$, below the scan. The $q_2q_3$ threshold at $\sqrt{s_{3,\mathrm{th}}}=m_2+m_3=0.986~\mathrm{GeV}$ belongs to the fixed $s_3$ channel. These reduced two-propagator Landau branch points are not crossed here; none of the listed coefficient zeros is a normal-threshold opening.

  Along this path, the Landau polynomial and the two linear factors become
  \begin{align}
    \mathcal{P}&=m_2^2(s_1-s_{1*})(s_1-s_{1\dagger}),\qquad s_{1\dagger}=2s_2+\frac{(m_1^2+m_2^2-s_2)s_{3*}}{m_2^2}-s_{1*},\label{eq:landau-scan-factors}\\
    \Xi&=s_1-r_{\Xi},\qquad r_{\Xi}=s_{3*}+2d-s_2,\qquad \Omega=s_1-r_{\Omega},\qquad r_{\Omega}=2s_{3*}+d-m_2^2-s_2.\label{eq:linear-scan-roots}
  \end{align}
  At $\sqrt{s_1}=1.42~\mathrm{GeV}$, the simple zeros of $\mathcal{P}$ and $H$ give simple zeros of $f_{C2}^{2},f_{C6}^{1}$ but a double zero of $f_{C12}^{1}$. The latter is a coincidence of numerator factors, not two Landau branch points. Their second common zero at $\sqrt{s_{1\dagger}}\simeq1.449776~\mathrm{GeV}$ fails the Coleman--Norton condition and produces no physical TS enhancement.

  The first component of $f_{C2}$ also shares the reference-point zero:
  \begin{equation}
    f_{C2}^{1}=\frac{s_1-s_{1*}}{G^2}\left\{m_2^2(s_1-s_{1\dagger})[G+2(s_1-s_{1*})(V+Z)]-\mathcal{K}\right\}.\label{eq:fC21-scan-factor}
  \end{equation}
  The cubic in braces has one root in the interval, at $\sqrt{s_1}\simeq1.430925~\mathrm{GeV}$, distinct from the second root of $f_{C2}^{2}$. The linear factors give
  \begin{equation}
    f_{C10}^{1}=s_1-r_{\Omega},\qquad f_{C11}^{1}=\frac{s_{3*}}{G}(s_1-s_{1*})(s_1-r_{\Xi})(s_1-r_{\Omega}).\label{eq:fC1011-scan-factors}
  \end{equation}
  The three roots of $f_{C11}^{1}$ arise from $H,\Xi,\Omega$; the last is shared with $f_{C10}^{1}$. Neither the $\Xi$ nor the $\Omega$ root is a Landau root for these inputs.

  After cancelling common numerator and denominator factors, all scanned coefficients can be written as
  \begin{equation}
    \left.f_{Ci}^{k}\right|_{\mathrm{scan}}=\frac{R_{ik}(s_1)}{D_{ik}(s_1)}\prod_a(s_1-r_{ik,a})^{n_{ik,a}},\qquad 1.4~\mathrm{GeV}\leq\sqrt{s_1}\leq1.5~\mathrm{GeV}.\label{eq:remaining-scan-factorization}
  \end{equation}
  Here $D_{ik}$ is the reduced denominator and $R_{ik}$ has no zeros in the interval. All roots $r_{ik,a}$, their defining conditions, and multiplicities are summarized in \cref{tab:remaining-C0-zeros}, with energies rounded to the displayed precision.

  \begin{table}[!ht]
    \centering
    \caption{Zeros of all seventeen physical-mass $C_0$ coefficients along the fixed-$s_3$, fixed-$s_{23}$ scan. Multiplicities follow the order of the listed energies. The last row groups the components with no real zero in the plotted interval.}
    \begin{ruledtabular}\label{tab:remaining-C0-zeros}
      \begin{tabular}{llll}
        Coefficient & Numerator condition & $\sqrt{s_1}$ (GeV) & Multiplicity\\\hline
        $f_{C2}^{1}$ & $[G+2H(V+Z)]\mathcal{P}-H\mathcal{K}=0$ & $1.420000,\ 1.430925$ & $1,\ 1$\\
        $f_{C2}^{2}$ & $\mathcal{P}=0$ & $1.420000,\ 1.449776$ & $1,\ 1$\\
        $f_{C3}^{1}$ & $[G+H(V+Z)]\mathcal{K}-8Hs_1Z\mathcal{P}=0$ & $1.434573$ & $1$\\
        $f_{C3}^{2}$ & $\mathcal{K}=0$ & $1.434629$ & $1$\\
        $f_{C6}^{1}$ & $\mathcal{P}=0$ & $1.420000,\ 1.449776$ & $1,\ 1$\\
        $f_{C7}^{1}$ & $Z\Xi\Omega-2\mathcal{P}=0$ & $1.436134$ & $1$\\
        $f_{C8}^{1}$ & $\mathcal{P}-2G\Omega=0$ & $1.496597$ & $1$\\
        $f_{C10}^{1}$ & $\Omega=0$ & $1.495754$ & $1$\\
        $f_{C11}^{1}$ & $H\Xi\Omega=0$ & $1.420000,\ 1.434329,\ 1.495754$ & $1,\ 1,\ 1$\\
        $f_{C12}^{1}$ & $H\mathcal{P}=0$ & $1.420000,\ 1.449776$ & $2,\ 1$\\
        \multicolumn{2}{l}{$f_{C1}^{1},f_{C4}^{1},f_{C5}^{1},f_{C8}^{2,3},f_{C9}^{1,2}$: no zero in the interval} & --- & ---
      \end{tabular}
    \end{ruledtabular}
  \end{table}

  The other roots in \cref{tab:remaining-C0-zeros} are numerator cancellations, not new singularities. The $f_{C3}^{1,2}$ roots differ by approximately $0.0557~\mathrm{MeV}$ despite appearing coincident in \cref{fig:4-3}. Likewise, the $f_{C8}^{1}$ root differs from the nearby $\Omega$ root, since $f_{C8}^{1}|_{\Omega=0}=\mathcal{P}/(2G)\neq0$. The table also identifies the components with no root in the interval.

  Since $s_1-r=(E-\sqrt r)(E+\sqrt r)$ with $E=\sqrt{s_1}>0$, a root of multiplicity $n$ gives $|f_{Ci}^{k}|\propto|E-E_0|^n$ locally. Finite sampling on a logarithmic plot turns an exact zero into a finite-depth minimum, whose depth has no independent physical significance.

  The $B_{03}$ scan in \cref{fig:B0coeff} distinguishes kinematic zeros from subtraction cancellations. The mass-independent coefficients in \cref{fig:5-1,fig:5-4,fig:5-9,fig:5-10} remain nonzero, but their complete bubble products vanish under subtraction of the missing propagator mass, explaining the zero $b_i^{3k}$ entries in \cref{tab:relativeloop}. By contrast, $f_{B11}^{31}$ and $f_{B12}^{31}$ vanish at the reference point through $H=0$.

  In \cref{fig:B0C0}, $\mathbf{B}_{03}$ varies smoothly, whereas $\mathbf{C}_0$ is enhanced at the physical TS near $1.42~\mathrm{GeV}$ but not at the second Landau root. Their different mass dimensions preclude interpreting the plotted magnitudes as relative amplitude weights.

  Coefficient zeros likewise do not determine zeros of the complete regulated amplitude. Individual component zeros can depend on the Lorentz basis, and all contributions must be combined coherently. Provided regulator-induced singularities remain outside the physical TS neighborhood, shifted-mass terms can leave a regular contribution without restoring a missing leading logarithm. The normal-threshold strength instead depends on the coherent B$_0$--C$_0$ combination in \cref{eq:normal-threshold-amplitude}; the cancellations in \cref{eq:threshold-cancellation} remove its leading square root, not all threshold dependence.

  These distinctions support the preceding lineshape classification: the class-(I) coefficients remain nonzero at the reference point; those in \cref{fig:4-3,fig:4-7} also remain nonzero, so their weaker peaks require interference rather than removal of the Landau logarithm. The midpoint-dependent zeros in \cref{fig:4-2,fig:4-11} differ from the generic Landau factors in \cref{fig:4-6,fig:4-12}. The full $s_3$-dependent amplitudes, not the coefficient minima alone, determine the spectra.

\section{Summary}\label{sec:summary}
In this work, we have systematically investigated how the dynamical couplings at the three vertices of a triangle diagram affect the visibility and lineshape of the triangle singularity. We enumerated twelve combinations of external and internal spin-parity quantum numbers, covering seven types of effective vertices: $PPS$, $VPP$, $VVP$, $VPA$, $VVV$, $VVA$, and $VVS$. Using the Passarino--Veltman reduction, we expressed each loop amplitude as a linear combination of the scalar one-loop integrals $A_{0}$, $B_{0}$, and $C_{0}$, with the corresponding Lorentz-scalar coefficients encoding the dynamical information of the vertices under consideration.

For the numerical comparison, we used the masses and kinematics of the $\eta(1405)\to\pi^{+}\pi^{-}\pi^{0}$ system with a $K^{*}\bar{K}K$ loop, together with the $f_{0}(980)$ resonance parameters, as a common benchmark for the twelve spin-parity assignments. We evaluated the relative weights $a_{i}^{jk}$ and $b_{i}^{jk}$ at the Landau singularity. The twelve cases can be classified into three categories. In the first, the $C_{0}$ contribution dominates and produces a pronounced narrow TS peak. In the second, the bubble contributions are more competitive, and their interference with the triangle term can reduce the prominence of the peak. In the third, the physical-mass coefficient of the leading Landau logarithm vanishes as a consequence of the vertex dynamics. The cusp-like spectrum is then shaped by the remaining ordinary-threshold dependence, smooth background, and coherent interference, without requiring an explicit $B_{0}$ term to become dominant. In particular, the leading normal-threshold square-root terms cancel in the complete amplitudes for cases $6$ and $12$, so their threshold behavior must be determined beyond that order.

These results demonstrate that the Landau condition determines the kinematic location at which a triangle singularity may occur, but does not by itself determine its prominence or detailed lineshape in the physical spectrum. The partial-wave and tensor structures of the vertices also play a crucial role. Within the effective-Lagrangian and intermediate-meson-loop framework adopted in this work, comparing the relative contributions of $B_{0}$ and $C_{0}$ allows us to analyze how the vertex dynamics affects TS interpretations of charmoniumlike and pentaquark candidates. Future TS analyses should therefore examine not only the Landau condition but also the relative magnitudes of the relevant scalar integrals and their coefficients. More precise measurements of the corresponding lineshapes at BESIII, LHCb, and Belle~II will help distinguish triangle-singularity effects from contributions associated with genuine resonances.

\acknowledgments
This work is supported by the National Natural Science Foundation of China (Grant No. 12235018).

\appendix
\linespread{1}
\section{Coefficients in the scalar-integral decomposition}\label{appendix:A}
In this appendix, we set $M_{k_1}=M_{k_2}=M_{f}$ and $m_3=m_2$. The latter equality implies
\begin{equation}
  \mathbf{A}_{02}=\mathbf{A}_{03}\,.
\end{equation}
For convenience, we use $s_1=p_1^{2}$, $s_2=p_2^{2}$, $s_3=(k_1+k_2)^{2}$, and $s_{23}=(k_2+p_2)^{2}$.

\begin{table}[!ht]
  \centering
  \caption{Coefficients of scalar loop integrals in the decomposition of $i \mathcal{M}_{1}$}
  \begin{ruledtabular}    \label{tab:coeff1}
    \begin{tabular}{cc}
      Coefficient & Expression\\\hline
      \multirow{1}{0.05\textwidth}{$A_{1}^{1}$}&\multirow{1}{0.9\textwidth}{0}\\
      \multirow{1}{0.05\textwidth}{$A_{1}^{2}$}&\multirow{1}{0.9\textwidth}{0}\\
      \multirow{1}{0.05\textwidth}{$A_{1}^{3}$}&\multirow{1}{0.9\textwidth}{0}\\
      \multirow{1}{0.05\textwidth}{$B_{1}^{1}$}&\multirow{1}{0.9\textwidth}{$\varepsilon_{1i}\cdot p_{2}(s_2-s_1-s_3)/\lambda(s_1,s_2,s_3)$}\\
      \multirow{1}{0.05\textwidth}{$B_{1}^{2}$}&\multirow{1}{0.9\textwidth}{$\varepsilon_{1i}\cdot p_{2}(s_3-s_1-s_2)/\lambda(s_1,s_2,s_3)$}\\
      \multirow{1}{0.05\textwidth}{$B_{1}^{3}$}&\multirow{1}{0.9\textwidth}{$2\varepsilon_{1i}\cdot p_{2}s_1/\lambda(s_1,s_2,s_3) $}\\
      \multirow{1}{0.05\textwidth}{$C_{1}$}&\multirow{1}{0.9\textwidth}{$\varepsilon_{1i}\cdot p_{2}(s_3(3s_1+4s_2-m_1^{2}+m_3^{2})-(s_1-s_2)(m_1^{2}-m_3^{2}+s_1-2s_2)-2s_3^{2})/\lambda(s_1,s_2,s_3)$}\\
    \end{tabular}
  \end{ruledtabular}
\end{table}

\begin{table}[!ht]
  \centering
  \caption{Coefficients of scalar loop integrals in the decomposition of $i \mathcal{M}_{2}$}
  \begin{ruledtabular}    \label{tab:coeff2}
    \begin{tabular}{cc}
      Coefficient & Expression\\\hline
      \multirow{2}{0.05\textwidth}{ $A_{2}^{1}$}&\multirow{2}{0.9\textwidth}{$\displaystyle -\frac{1}{2 s_2 (s_1^2-2 (s_2+s_3) s_1+(s_2-s_3)^2)}(-s_1+s_2+s_3) (2 M_f^2+s_1+s_2-s_3-2 s_{23}) (p_2 \cdot\varepsilon_{1i}  )$}\vspace{10pt}\\[1pt]
      \multirow{2}{0.05\textwidth}{ $A_{2}^{2}$}&\multirow{2}{0.9\textwidth}{$\displaystyle \frac{1}{2 s_2 (s_1^2-2 (s_2+s_3) s_1+(s_2-s_3)^2)}(-s_1+s_2+s_3) (2 M_f^2+s_1+s_2-s_3-2 s_{23}) (p_2 \cdot\varepsilon_{1i}  )$}\vspace{10pt}\\[1pt]
      \multirow{1}{0.05\textwidth}{ $A_{2}^{3}$}&$\setminus$\\[1pt]
      \multirow{6}{0.05\textwidth}{ $B_{2}^{1}$}&\multirow{6}{0.9\textwidth}{$\displaystyle \frac{1}{(s_1^2-2 (s_2+s_3) s_1+(s_2-s_3)^2)^2}s_3 ((p_2 \cdot\varepsilon_{1i}  ) (s_3^2 (9 M_f^2-2 m_1^2+2 m_2^2+9 s_1+12 s_2-9 s_{23})-2 s_3 (m_1^2 (-3 M_f^2+s_1-2 s_2+3 s_{23})+m_2^2 (3 M_f^2-s_1+2 s_2-3 s_{23})+6 s_1 M_f^2+9 s_2 M_f^2+3 s_1^2+6 s_2^2+4 s_1 s_2-6 s_1 s_{23}-9 s_2 s_{23})+(s_1-s_2) (2 m_1^2 (3 M_f^2+2 s_1+s_2-3 s_{23})-2 m_2^2 (3 M_f^2+2 s_1+s_2-3 s_{23})-9 s_2 M_f^2+3 s_1 M_f^2+s_1^2-4 s_2^2-3 s_1 s_2-3 s_1 s_{23}+9 s_2 s_{23})-4 s_3^3)+(s_1^2-2 (s_2+s_3) s_1+(s_2-s_3)^2) (2 m_1^2-2 m_2^2-s_1-s_2+s_3) (k_1 \cdot\varepsilon_{1i}  ))$}\vspace{50pt}\\[1pt]
      \multirow{10}{0.05\textwidth}{$B_{2}^{2}$}&\multirow{10}{0.9\textwidth}{$\displaystyle \frac{1}{2 s_2 (s_1^2-2 (s_2+s_3) s_1+(s_2-s_3)^2)^2}\{2 s_2 (s_1^2-2 (s_2+s_3) s_1+(s_2-s_3)^2) ((m_1^2-m_2^2-s_2) (s_1-s_2)-(m_1^2-m_2^2+s_2) s_3) (k_1 \cdot  \varepsilon_{1i})+(-((s_3^4-2 (M_f^2+2 s_1+3 s_2-s_{23}) s_3^3+2 (3 (s_1+2 s_2) M_f^2+3 s_1^2+4 s_2^2+4 s_1 s_2-3 (s_1+2 s_2) s_{23}) s_3^2-2 (2 s_1^3-(s_2+3 s_{23}) s_1^2+2 s_2 (3 s_2-s_{23}) s_1+M_f^2 (3 s_1^2+2 s_2 s_1+3 s_2^2)+s_2^2 (s_2-3 s_{23})) s_3+(s_1-s_2)^2 (2 (s_1-2 s_2) M_f^2+s_1^2-2 s_1 (s_2+s_{23})-s_2 (s_2-4 s_{23}))) m_1^2)+s_2 (5 s_3^4-2 (5 M_f^2+9 s_1+6 s_2-5 s_{23}) s_3^3+2 ((13 s_1+6 s_2) M_f^2+12 s_1^2+3 s_2 (s_2-2 s_{23})+s_1 (10 s_2-13 s_{23})) s_3^2-2 (7 s_1^3-11 s_{23} s_1^2+3 s_2^2 s_1+s_2^2 (3 s_{23}-2 s_2)+M_f^2 (11 s_1^2-3 s_2^2)) s_3+(s_1-s_2)^2 ((6 s_1-8 s_2) M_f^2+3 s_1^2+s_2 (8 s_{23}-3 s_2)-2 s_1 (s_2+3 s_{23})))+m_2^2 (s_3^4-2 (M_f^2+2 s_1+3 s_2-s_{23}) s_3^3+2 (3 (s_1+2 s_2) M_f^2+3 s_1^2+4 s_2^2+4 s_1 s_2-3 (s_1+2 s_2) s_{23}) s_3^2-2 (2 s_1^3-(s_2+3 s_{23}) s_1^2+2 s_2 (3 s_2-s_{23}) s_1+M_f^2 (3 s_1^2+2 s_2 s_1+3 s_2^2)+s_2^2 (s_2-3 s_{23})) s_3+(s_1-s_2)^2 (2 (s_1-2 s_2) M_f^2+s_1^2-2 s_1 (s_2+s_{23})-s_2 (s_2-4 s_{23})))) (p_2 \cdot  \varepsilon_{1i})\}$}\vspace{90pt}\\[1pt]
      \multirow{7}{0.05\textwidth}{ $B_{2}^{3}$}&\multirow{7}{0.9\textwidth}{$\displaystyle\frac{1}{(s_1^2-2 (s_2+s_3) s_1+(s_2-s_3)^2)^2}\{(p_2 \cdot\varepsilon_{1i}  ) (s_3^2 (m_2^2 (M_f^2-5 s_1-s_{23})-(s_1-12 s_2) M_f^2-3 s_1^2+12 s_2^2+s_1 s_{23}-12 s_2 s_{23})+2 s_3 (m_2^2 ((5 s_1-s_2) M_f^2+2 s_1^2+2 s_1 s_2-5 s_1 s_{23}+s_2 s_{23})+(4 s_1^2+6 s_2 s_1-6 s_2^2) M_f^2+2 s_1^3+3 s_2 s_1^2-4 s_{23} s_1^2+3 s_2^2 s_1-6 s_2 s_{23} s_1-4 s_2^3+6 s_2^2 s_{23})+(s_1-s_2)^2 (m_2^2 (M_f^2+s_1-s_{23})+(4 s_2-3 s_1) M_f^2-s_1^2+2 s_2^2+3 s_1 s_{23}-4 s_2 s_{23})+m_1^2 (-((s_1-s_2)^2 (M_f^2+s_1-s_{23}))+s_3^2 (-M_f^2+5 s_1+s_{23})-2 s_3 ((5 s_1-s_2) M_f^2+2 s_1^2+2 s_1 s_2-5 s_1 s_{23}+s_2 s_{23}))-2 s_3^3 (2 M_f^2+s_1+4 s_2-2 s_{23})+2 s_3^4)-(s_1^2-2 (s_2+s_3) s_1+(s_2-s_3)^2) (m_1^2 (s_1-s_2+s_3)-m_2^2 (s_1-s_2+s_3)+s_1 (-s_1+s_2+s_3)) (k_1 \cdot\varepsilon_{1i}  )\}$}\vspace{60pt}\\[1pt]
    \end{tabular}
  \end{ruledtabular}
\end{table}

\begin{table}[!ht]
  \centering
  \ContinuedFloat
  \caption{(Continued)}
  \begin{ruledtabular}
    \begin{tabular}{cc}
      Coefficient & Expression\\\hline
      \multirow{11}{0.05\textwidth}{$C_{2}$}&\multirow{11}{0.9\textwidth}{$\displaystyle\frac{1}{(s_1^2-2 (s_2+s_3) s_1+(s_2-s_3)^2)^2}\{2 (s_1^2-2 (s_2+s_3) s_1+(s_2-s_3)^2) (s_3 m_2^4+((s_1-s_2)^2-(2 m_1^2+s_1+s_2) s_3) m_2^2+(m_1^4-(s_1+s_2-s_3) m_1^2+s_1 s_2) s_3) (k_1 \cdot  \varepsilon_{1i})+(-2 s_3 (s_3^2+(-3 M_f^2+s_1-2 s_2+3 s_{23}) s_3-(s_1-s_2) (3 M_f^2+2 s_1+s_2-3 s_{23})) m_2^4+(5 s_3^4+(4 m_1^2-10 M_f^2-11 s_1-15 s_2+10 s_{23}) s_3^3+(4 (-3 M_f^2+s_1-2 s_2+3 s_{23}) m_1^2+9 s_1^2+15 s_2^2+12 s_1 s_2+2 M_f^2 (5 s_1+9 s_2)-10 s_1 s_{23}-18 s_2 s_{23}) s_3^2-(s_1-s_2) (4 (3 M_f^2+2 s_1+s_2-3 s_{23}) m_1^2+5 s_1^2-5 s_2^2-4 s_1 s_2+2 M_f^2 (s_1-3 s_2)-2 s_1 s_{23}+6 s_2 s_{23}) s_3+2 (s_1-s_2)^3 (M_f^2+s_1-s_{23})) m_2^2-s_3 (2 (s_3^2+(-3 M_f^2+s_1-2 s_2+3 s_{23}) s_3-(s_1-s_2) (3 M_f^2+2 s_1+s_2-3 s_{23})) m_1^4+(s_1^3+(5 s_2+3 s_3) s_1^2+(12 (s_2+s_3) M_f^2-9 s_3^2-s_2 (s_2+12 s_{23})+2 s_3 (5 s_2-6 s_{23})) s_1-(s_2-s_3)^2 (12 M_f^2+5 s_2-5 s_3-12 s_{23})) m_1^2+2 s_3^4+2 M_f^2 (s_1^3-(3 s_2+4 s_3) s_1^2+5 (s_3-s_2) s_3 s_1+2 (s_2-s_3)^3)+(-5 s_1^3-4 (s_2-2 s_{23}) s_1^2+s_2 (10 s_{23}-9 s_2) s_1+4 s_2^2 (3 s_{23}-2 s_2)) s_3+s_3^2 (9 s_1^2+5 (3 s_2-2 s_{23}) s_1+12 s_2 (s_2-s_{23}))+s_3^3 (-7 s_1-8 s_2+4 s_{23})+(s_1-s_2) (s_1^3-2 (s_2+s_{23}) s_1^2+s_2 (4 s_{23}-3 s_2) s_1-2 s_2^2 (s_2-2 s_{23})))) (p_2 \cdot  \varepsilon_{1i})\}$}\vspace{100pt} \\[5pt]
    \end{tabular}
  \end{ruledtabular}
\end{table}

\begin{table}[!ht]
  \centering
  \caption{Coefficients of scalar loop integrals in the decomposition of $i \mathcal{M}_{3}$}
  \begin{ruledtabular}   \label{tab:coeff3}
    \begin{tabular}{cc}
      Coefficient & Expression\\\hline
      \multirow{5}{0.05\textwidth}{  $A_{3}^{1}$}&\multirow{5}{0.9\textwidth}{$\displaystyle  \frac{\varepsilon_{1i}\cdot k_1 (-s_1+s_2+s_3)}{4 s_2}+\frac{\varepsilon_{1i}\cdot p_2 (-s_1+s_2+s_3) (s_1^2+M_f^2 (s_1-s_2+s_3)+(s_2-s_3)
      s_{23}-s_1 (s_2+s_3+s_{23}))}{4 s_2 (s_1^2+(s_2-s_3)^2-2 s_1 (s_2+s_3))}$}\vspace{40pt}\\[5pt]
      \multirow{5}{0.05\textwidth}{$A_{3}^{2}$}&\multirow{5}{0.9\textwidth}{$\displaystyle  -\frac{\varepsilon_{1i}\cdot k_1 (-s_1+s_2+s_3)}{4 s_2}-\frac{\varepsilon_{1i}\cdot p_2 (-s_1+s_2+s_3) (s_1^2+M_f^2 (s_1-s_2+s_3)+(s_2-s_3)
      s_{23}-s_1 (s_2+s_3+s_{23}))}{4 s_2 (s_1^2+(s_2-s_3)^2-2 s_1 (s_2+s_3))} $}\vspace{40pt}\\[1pt]
      $A_{3}^{3}$&$\setminus$\\
      \multirow{11}{0.05\textwidth}{$B_{3}^{1}$}&\multirow{11}{0.9\textwidth}{$\displaystyle \frac{1}{4 (s_1^2+(s_2-s_3)^2-2 s_1 (s_2+s_3))}\varepsilon_{1i}\cdot k_1 s_3 (-((2 m_1^2-2 m_3^2+5 s_1-7 s_2) (s_1-s_2))+2 (-m_1^2+m_3^2+6 s_1+7 s_2)
        s_3-7 s_3^2)+\frac{1}{4 (s_1^2+(s_2-s_3)^2-2 s_1 (s_2+s_3))^2}\varepsilon_{1i}\cdot p_2 s_3 (-s_3^3 (7 M_f^2-11
          s_1-7 s_{23})+(s_1-s_2)^2 (2 m_3^2-5 s_1+7 s_2) (M_f^2+s_1-s_{23})-s_3 (-21 s_1^3+2 s_1^2 s_2+3 s_1
            s_2^2+M_f^2 (-15 s_1^2-22 s_1 s_2+21 s_2^2)+15 s_1^2 s_{23}+22 s_1 s_2 s_{23}-21 s_2^2 s_{23}-4 m_3^2 (2 s_1^2+M_f^2 (5 s_1-s_2)+2
          s_1 s_2-5 s_1 s_{23}+s_2 s_{23}))-2 m_1^2 (s_3^2 (M_f^2-5 s_1-s_{23})+(s_1-s_2)^2 (M_f^2+s_1-s_{23})+2
          s_3 (2 s_1^2+M_f^2 (5 s_1-s_2)+2 s_1 s_2-5 s_1 s_{23}+s_2 s_{23}))-s_3^2 (2 m_3^2 (-M_f^2+5 s_1+s_{23})+3
      (9 s_1^2+M_f^2 (s_1-7 s_2)+5 s_1 s_2-s_1 s_{23}+7 s_2 s_{23})))$}\vspace{100pt}\\[1pt]
      \multirow{18}{0.05\textwidth}{$B_{3}^{2}$}&\multirow{18}{0.9\textwidth}{$\displaystyle\frac{1}{4 s_2 (s_1^2+(s_2-s_3)^2-2
        s_1 (s_2+s_3))}\varepsilon_{1i}\cdot k_1 ((s_1-s_2)^2 (m_1^2 s_1-m_3^2 s_1+s_2 (-3 s_1+2 s_2))+(s_2 (11 s_1^2-4 s_1
          s_2+s_2^2)-m_1^2 (3 s_1^2+2 s_1 s_2+3 s_2^2)+m_3^2 (3 s_1^2+2 s_1 s_2+3 s_2^2)) s_3+(m_1^2 (3 s_1+4 s_2)-m_3^2
        (3 s_1+4 s_2)-s_2 (13 s_1+8 s_2)) s_3^2+(-m_1^2+m_3^2+5 s_2) s_3^3)+\frac{1}{4 s_2 (s_1^2+(s_2-s_3)^2-2
        s_1 (s_2+s_3))^2}\varepsilon_{1i}\cdot p_2 (m_1^2 (s_1 (s_1-s_2)^3 (M_f^2+s_1-s_{23})+s_3^4 (-M_f^2+s_1+s_{23})+s_3^3
            (-4 s_1^2-7 s_1 s_2+M_f^2 (2 s_1+5 s_2)-2 s_1 s_{23}-5 s_2 s_{23})-(s_1-s_2) s_3 (4 s_1^3+M_f^2 (2 s_1^2+11
            s_1 s_2+3 s_2^2)+s_1 s_2 (7 s_2-11 s_{23})+s_1^2 (5 s_2-2 s_{23})-3 s_2^2 s_{23})+s_3^2 (6 s_1^3+11 s_1^2
          s_2+7 s_2^2 (-M_f^2+s_{23})-s_1 s_2 (-7 M_f^2+s_2+7 s_{23})))+m_3^2 (-s_1 (s_1-s_2)^3 (M_f^2+s_1-s_{23})-s_3^4
            (-M_f^2+s_1+s_{23})+s_3^3 (4 s_1^2+7 s_1 s_2-M_f^2 (2 s_1+5 s_2)+2 s_1 s_{23}+5 s_2 s_{23})+(s_1-s_2)
            s_3 (4 s_1^3+M_f^2 (2 s_1^2+11 s_1 s_2+3 s_2^2)+s_1 s_2 (7 s_2-11 s_{23})+s_1^2 (5 s_2-2 s_{23})-3 s_2^2 s_{23})+s_3^2
          (-6 s_1^3-11 s_1^2 s_2+7 s_2^2 (M_f^2-s_{23})+s_1 s_2 (-7 M_f^2+s_2+7 s_{23})))+s_2 (-((3 s_1-2
            s_2) (s_1-s_2)^3 (M_f^2+s_1-s_{23}))-5 s_3^4 (-M_f^2+s_1+s_{23})+s_3^3 (18 s_1^2-s_1 s_2-M_f^2
            (8 s_1+13 s_2)+8 s_1 s_{23}+13 s_2 s_{23})+s_3^2 (-24 s_1^3+M_f^2 (-2 s_1^2+9 s_1 s_2+9 s_2^2)+3 s_1 s_2 (5
            s_2-3 s_{23})-9 s_2^2 s_{23}+s_1^2 (9 s_2+2 s_{23}))+(s_1-s_2) s_3 (14 s_1^3+M_f^2 (8 s_1^2+9 s_1 s_2-s_2^2)+s_1
      s_2 (7 s_2-9 s_{23})+s_2^2 s_{23}-s_1^2 (5 s_2+8 s_{23}))))  $}\vspace{175pt}\\[1pt]

    \end{tabular}
  \end{ruledtabular}
\end{table}
\begin{table}[!ht]
  \centering
  \ContinuedFloat
  \caption{(Continued)}
  \begin{ruledtabular}
    \begin{tabular}{cc}
      Coefficient & Expression\\\hline
      \multirow{14}{0.05\textwidth}{ $B_{3}^{3}$}&\multirow{14}{0.9\textwidth}{$\displaystyle    \frac{1}{4 (s_1^2+(s_2-s_3)^2-2 s_1 (s_2+s_3))}\varepsilon_{1i}\cdot k_1 ((-m_1^2+m_3^2+5 s_1-4 s_2) (s_1-s_2)^2-2 (4 s_1^2+6 s_1 s_2-6 s_2^2-m_1^2 (3 s_1+s_2)+m_3^2
        (3 s_1+s_2)) s_3-(m_1^2-m_3^2+s_1+12 s_2) s_3^2+4 s_3^3)+\frac{1}{4 (s_1^2+(s_2-s_3)^2-2 s_1 (s_2+s_3))^2}\varepsilon_{1i}
        \cdot p_2 ((5 s_1-4 s_2) (s_1-s_2)^3 (M_f^2+s_1-s_{23})-4 s_3^4 (-M_f^2+s_1+s_{23})+s_3^3 (s_1^2+M_f^2
          (3 s_1-16 s_2)+8 s_1 s_2-3 s_1 s_{23}+16 s_2 s_{23})+s_3^2 (s_1^2 (15 s_1+s_2)-M_f^2 (s_1-s_2) (s_1+24
          s_2)+(s_1-s_2) (s_1+24 s_2) s_{23})-(s_1-s_2) s_3 (17 s_1^3+M_f^2 (11 s_1^2+21 s_1 s_2-16
          s_2^2)+s_1^2 (7 s_2-11 s_{23})+16 s_2^2 s_{23}-s_1 s_2 (8 s_2+21 s_{23}))+m_1^2 (-(s_1-s_2)^3
            (M_f^2+s_1-s_{23})+s_3^3 (-M_f^2-3 s_1+s_{23})+(s_1-s_2) s_3 (11 s_1^2+5 s_1 s_2+M_f^2 (13 s_1+3 s_2)-13
          s_1 s_{23}-3 s_2 s_{23})+s_3^2 (-7 s_1^2+7 s_1 s_2+M_f^2 (13 s_1+3 s_2)-13 s_1 s_{23}-3 s_2 s_{23}))+m_3^2 (M_f^2
            (s_1-s_2+s_3) (s_1^2+(s_2-s_3)^2-2 s_1 (s_2+7 s_3))+(s_1-s_2)^3 (s_1-s_{23})+s_3^3
            (3 s_1-s_{23})-(s_1-s_2) s_3 (11 s_1^2+5 s_1 s_2-13 s_1 s_{23}-3 s_2 s_{23})+s_3^2 (7 s_1^2-7 s_1 s_2+13 s_1
      s_{23}+3 s_2 s_{23}))) $}\vspace{140pt}\\[1pt]
      \multirow{20}{0.05\textwidth}{$C_{3}$}&\multirow{20}{0.9\textwidth}{$\displaystyle\frac{1}{2 (s_1^2+(s_2-s_3)^2-2 s_1 (s_2+s_3))}\varepsilon_{1i}\cdot k_1 (m_3^2 (s_1-s_2)^3-(s_1-s_2) (m_1^4+m_3^4-s_1^2-m_3^2 (s_1-7 s_2)+2 s_2^2-2 m_1^2
          (m_3^2-s_1+2 s_2)) s_3-(m_1^4+m_3^4+4 s_1^2+3 s_1 s_2+6 s_2^2-2 m_1^2 (m_3^2+3 s_1+4 s_2)+m_3^2 (7 s_1+11 s_2))
        s_3^2-(4 m_1^2-5 (m_3^2+s_1)-6 s_2) s_3^3-2 s_3^4)+\frac{1}{2 (s_1^2+(s_2-s_3)^2-2 s_1 (s_2+s_3))^2}\varepsilon_{1i}
        \cdot p_2 (-m_3^4 s_3 (s_3^2 (M_f^2-5 s_1-s_{23})+(s_1-s_2)^2 (M_f^2+s_1-s_{23})+2 s_3 (2 s_1^2+M_f^2
          (5 s_1-s_2)+2 s_1 s_2-5 s_1 s_{23}+s_2 s_{23}))+m_3^2 ((s_1-s_2)^4 (M_f^2+s_1-s_{23})-5 s_3^4
            (-M_f^2+s_1+s_{23})+2 (s_1-s_2)^2 s_3 (-2 s_1^2-5 s_1 s_2-M_f^2 (3 s_1+4 s_2)+m_1^2 (M_f^2+s_1-s_{23})+3
            s_1 s_{23}+4 s_2 s_{23})+2 s_3^3 (s_1 (4 s_1+s_2)-M_f^2 (s_1+8 s_2)+m_1^2 (M_f^2-5 s_1-s_{23})+(s_1+8
            s_2) s_{23})+2 s_3^2 (2 s_1 s_2 (s_1+3 s_2)+M_f^2 (s_1^2-2 s_1 s_2+9 s_2^2)-(s_1^2-2 s_1 s_2+9 s_2^2)
          s_{23}+2 m_1^2 (2 s_1^2+M_f^2 (5 s_1-s_2)+2 s_1 s_2-5 s_1 s_{23}+s_2 s_{23})))-s_3 ((s_1-s_2)^2
            (m_1^4-s_1^2+2 m_1^2 (s_1-2 s_2)+2 s_2^2) (M_f^2+s_1-s_{23})-2 s_3^4 (-M_f^2+s_1+s_{23})+s_3^3 (7
            s_1^2+4 s_1 s_2-M_f^2 (3 s_1+8 s_2)+4 m_1^2 (M_f^2-2 s_1-s_{23})+3 s_1 s_{23}+8 s_2 s_{23})+s_3^2 (-s_1^2 (9
              s_1+8 s_2)+M_f^2 (-s_1^2+2 s_1 s_2+12 s_2^2)+m_1^4 (M_f^2-5 s_1-s_{23})+(s_1^2-2 s_1 s_2-12 s_2^2) s_{23}+6
            m_1^2 (3 s_1^2+M_f^2 (s_1-2 s_2)+2 s_1 s_2-s_1 s_{23}+2 s_2 s_{23}))+s_3 (M_f^2 (3 s_1^3+8 s_1^2 s_2+5 s_1
              s_2^2-8 s_2^3)+s_1 (5 s_1^3+2 s_1^2 s_2+5 s_1 s_2^2-4 s_2^3)-(3 s_1^3+8 s_1^2 s_2+5 s_1 s_2^2-8 s_2^3) s_{23}+2 m_1^4
              (2 s_1^2+M_f^2 (5 s_1-s_2)+2 s_1 s_2-5 s_1 s_{23}+s_2 s_{23})-4 m_1^2 (3 s_1^3+M_f^2 (3 s_1^2+4 s_1 s_2-3 s_2^2)+s_1^2
      (s_2-3 s_{23})-4 s_1 s_2 s_{23}+3 s_2^2 s_{23}))))$}\vspace{200pt}\\[1pt]
    \end{tabular}
  \end{ruledtabular}
\end{table}

\begin{table}[!ht]
  \centering
  \caption{Coefficients of scalar loop integrals in the decomposition of $i \mathcal{M}_{4}$}
  \begin{ruledtabular}   \label{tab:coeff4}
    \begin{tabular}{cc}
      Coefficient & Expression\\\hline
      \multirow{1}{0.05\textwidth}{ $A_{4}^{1}$}&\multirow{1}{0.9\textwidth}{0}\\[1pt]
      \multirow{1}{0.05\textwidth}{ $A_{4}^{2}$}&\multirow{1}{0.9\textwidth}{0}\\[1pt]
      \multirow{1}{0.05\textwidth}{ $A_{4}^{3}$}&\multirow{1}{0.9\textwidth}{$\setminus$}\\[1pt]
      \multirow{2}{0.05\textwidth}{ $B_{4}^{1}$}&\multirow{2}{0.9\textwidth}{$\displaystyle  \frac{\varepsilon_{\mu \nu \rho \sigma}\varepsilon_{1i}^{\mu}\varepsilon_{1f}^{*\nu}p_{1}^{\rho}p_{2}^{\sigma}(s_1-s_2+s_3)}{s_1^2+(s_2-s_3)^2-2 s_1 (s_2+s_3)}$}\vspace{10pt}\\[1pt]
      \multirow{2}{0.05\textwidth}{ $B_{4}^{2}$}&\multirow{2}{0.9\textwidth}{$\displaystyle  \frac{\varepsilon_{\mu \nu \rho \sigma}\varepsilon_{1i}^{\mu}\varepsilon_{1f}^{*\nu}p_{1}^{\rho}p_{2}^{\sigma}  (s_1+s_2-s_3)}{s_1^2+(s_2-s_3)^2-2 s_1 (s_2+s_3)} $}\vspace{10pt}\\[1pt]
      \multirow{2}{0.05\textwidth}{ $B_{4}^{3}$}&\multirow{2}{0.9\textwidth}{$\displaystyle   -\frac{2 \varepsilon_{\mu \nu \rho \sigma}\varepsilon_{1i}^{\mu}\varepsilon_{1f}^{*\nu}p_{1}^{\rho}p_{2}^{\sigma}  s_1}{s_1^2+(s_2-s_3)^2-2 s_1 (s_2+s_3)} $}\vspace{10pt}\\[1pt]
      \multirow{2}{0.05\textwidth}{ $C_{4}$}&\multirow{2}{0.9\textwidth}{$\displaystyle  \frac{\varepsilon_{\mu \nu \rho \sigma}\varepsilon_{1i}^{\mu}\varepsilon_{1f}^{*\nu}p_{1}^{\rho}p_{2}^{\sigma}}{s_1^2+(s_2-s_3)^2-2
      s_1 (s_2+s_3)} ((m_1^2-m_2^2+s_1-2 s_2) (s_1-s_2)+(m_1^2-m_2^2-3 s_1-4 s_2) s_3+2 s_3^2) $}\vspace{10pt}\\[1pt]
    \end{tabular}
  \end{ruledtabular}
\end{table}

\begin{table}[!ht]
  \centering
  \caption{Coefficients of scalar loop integrals in the decomposition of $i \mathcal{M}_{5}$}
  \begin{ruledtabular}   \label{tab:coeff5}
    \begin{tabular}{cc}
      Coefficient & Expression\\\hline
      \multirow{2}{0.05\textwidth}{ $A_{5}^{1}$}&\multirow{2}{0.9\textwidth}{$\displaystyle \frac{\varepsilon_{1i}\cdot p_{2} \varepsilon_{1f}^{*}\cdot k_{2}}{s_2}$}\vspace{10pt}\\[1pt]
      \multirow{2}{0.05\textwidth}{ $A_{5}^{2}$}&\multirow{2}{0.9\textwidth}{$\displaystyle  -\frac{\varepsilon_{1i}\cdot p_{2} \varepsilon_{1f}^{*}\cdot k_{2}}{s_2} $}\vspace{10pt}\\[1pt]
      \multirow{1}{0.05\textwidth}{ $A_{5}^{3}$}&\multirow{1}{0.9\textwidth}{$\setminus$}\\[1pt]
      \multirow{2}{0.05\textwidth}{ $B_{5}^{1}$}&\multirow{2}{0.9\textwidth}{$\displaystyle  \frac{\varepsilon_{1i}\cdot p_{2} \varepsilon_{1f}^{*}\cdot k_{2} }{s_1^2+(s_2-s_3)^2-2
      s_1 (s_2+s_3)}(5 (s_1-s_2)^2+2 (s_1+s_2) s_3-7 s_3^2+2 m_1^2 (s_1-s_2+s_3)-2 m_2^2 (s_1-s_2+s_3)) $}\vspace{10pt}\\[1pt]
      \multirow{4}{0.05\textwidth}{ $B_{5}^{2}$}&\multirow{4}{0.9\textwidth}{$\displaystyle \frac{\varepsilon_{1i}\cdot p_{2} \varepsilon_{1f}^{*}\cdot k_{2}}{s_2 (s_1^2+(s_2-s_3)^2-2 s_1 (s_2+s_3))} (s_2 (5 s_1^2-7 s_2^2+2 s_1 (s_2-5 s_3)+2 s_2 s_3+5 s_3^2)+m_2^2 (s_1^2-s_2^2+s_3^2-2 s_1 (2
      s_2+s_3))+m_1^2 (-s_1^2+s_2^2-s_3^2+2 s_1 (2 s_2+s_3)))$}\vspace{30pt}\\[1pt]
      \multirow{2}{0.05\textwidth}{ $B_{5}^{3}$}&\multirow{2}{0.9\textwidth}{$\displaystyle  -\frac{4 \varepsilon_{1i}\cdot p_{2} \varepsilon_{1f}^{*}\cdot k_{2}}{s_1^2+(s_2-s_3)^2-2 s_1 (s_2+s_3)} (m_1^2 s_1-m_2^2 s_1+s_1^2+s_1 s_2-2 s_2^2+s_1 s_3+4 s_2 s_3-2 s_3^2)$}\vspace{10pt}\\[1pt]
      \multirow{4}{0.05\textwidth}{ $C_{5}$}&\multirow{4}{0.9\textwidth}{$\displaystyle  \frac{2 \varepsilon_{1i}\cdot p_{2} \varepsilon_{1f}^{*}\cdot k_{2}}{s_1^2+(s_2-s_3)^2-2 s_1 (s_2+s_3)} ((s_1-s_2) (m_1^4+(m_2^2+s_1)^2+m_1^2 (-2 m_2^2+2 s_1-3 s_2)-(m_2^2+3 s_1)
          s_2+2 s_2^2)+(m_1^4+m_2^4-4 s_1^2+m_1^2 (-2 m_2^2+s_1)-6 s_1 s_2+2 s_2^2-m_2^2 (9 s_1+8 s_2)) s_3+(-3
      m_1^2+7 m_2^2+5 s_1+2 s_2) s_3^2-2 s_3^3)$}\vspace{30pt}\\[1pt]
    \end{tabular}
  \end{ruledtabular}
\end{table}
\begin{table}[!ht]
  \centering
  \caption{Coefficients of scalar loop integrals in the decomposition of $i \mathcal{M}_{6}$}
  \begin{ruledtabular}   \label{tab:coeff6}
    \begin{tabular}{cc}
      Coefficient & Expression\\\hline
      \multirow{1}{0.05\textwidth}{ $A_{6}^{1}$}&\multirow{1}{0.9\textwidth}{$\displaystyle 0$}\\[1pt]
      \multirow{1}{0.05\textwidth}{ $A_{6}^{2}$}&\multirow{1}{0.9\textwidth}{$\displaystyle 0$}\\[1pt]
      \multirow{1}{0.05\textwidth}{ $A_{6}^{3}$}&\multirow{1}{0.9\textwidth}{$\setminus$}\\[1pt]
      \multirow{2}{0.05\textwidth}{ $B_{6}^{1}$}&\multirow{2}{0.9\textwidth}{$\displaystyle  \frac{2 \varepsilon_{\mu\nu\rho\sigma}\varepsilon_{1i}^{\mu}p_{2}^{\nu}k_{1}^{\rho}k_{2}^{\sigma}}{s_1^2+(s_2-s_3)^2-2 s_1 (s_2+s_3)} (-2 m_1^2+2 m_2^2+s_1+s_2-s_3) s_3$}\vspace{10pt}\\[1pt]
      \multirow{2}{0.05\textwidth}{ $B_{6}^{2}$}&\multirow{2}{0.9\textwidth}{$\displaystyle  \frac{2 \varepsilon_{\mu\nu\rho\sigma}\varepsilon_{1i}^{\mu}p_{2}^{\nu}k_{1}^{\rho}k_{2}^{\sigma} }{s_1^2+(s_2-s_3)^2-2
      s_1 (s_2+s_3)}((s_1-s_2) (-m_1^2+m_2^2+s_2)+(m_1^2-m_2^2+s_2) s_3) $}\vspace{10pt}\\[1pt]
      \multirow{2}{0.05\textwidth}{ $B_{6}^{3}$}&\multirow{2}{0.9\textwidth}{$\displaystyle  \frac{2 \varepsilon_{\mu\nu\rho\sigma}\varepsilon_{1i}^{\mu}p_{2}^{\nu}k_{1}^{\rho}k_{2}^{\sigma} }{s_1^2+(s_2-s_3)^2-2
      s_1 (s_2+s_3)}(m_1^2 (s_1-s_2+s_3)-m_2^2 (s_1-s_2+s_3)+s_1 (-s_1+s_2+s_3)) $}\vspace{10pt}\\[1pt]
      \multirow{2}{0.05\textwidth}{ $C_{6}$}&\multirow{2}{0.9\textwidth}{$\displaystyle \frac{4 \varepsilon_{\mu\nu\rho\sigma}\varepsilon_{1i}^{\mu}p_{2}^{\nu}k_{1}^{\rho}k_{2}^{\sigma} }{s_1^2+(s_2-s_3)^2-2 s_1 (s_2+s_3)}(-m_2^4 s_3-(m_1^4+s_1 s_2-m_1^2 (s_1+s_2-s_3)) s_3+m_2^2 (-(s_1-s_2)^2+(2
      m_1^2+s_1+s_2) s_3)) $}\vspace{10pt}\\[1pt]
    \end{tabular}
  \end{ruledtabular}
\end{table}
\begin{table}[!ht]
  \centering
  \caption{Coefficients of scalar loop integrals in the decomposition of $i \mathcal{M}_{7}$}
  \begin{ruledtabular}   \label{tab:coeff7}
    \begin{tabular}{cc}
      Coefficient & Expression\\\hline
      \multirow{2}{0.05\textwidth}{ $A_{7}^{1}$}&\multirow{2}{0.9\textwidth}{$\displaystyle -\frac{\varepsilon_{\mu\nu\rho\sigma}\varepsilon_{1i}^{\mu}p_{2}^{\nu}k_{1}^{\rho}k_{2}^{\sigma}  (s_1+s_2)}{s_1 s_2}$}\vspace{10pt}\\[1pt]
      \multirow{2}{0.05\textwidth}{ $A_{7}^{2}$}&\multirow{2}{0.9\textwidth}{$\displaystyle \frac{\varepsilon_{\mu\nu\rho\sigma}\varepsilon_{1i}^{\mu}p_{2}^{\nu}k_{1}^{\rho}k_{2}^{\sigma}  (s_1+s_2)}{s_1 s_2}$}\vspace{10pt}\\[1pt]
      \multirow{1}{0.05\textwidth}{ $A_{7}^{3}$}&\multirow{1}{0.9\textwidth}{$\setminus$}\\[1pt]
      \multirow{3}{0.05\textwidth}{ $B_{7}^{1}$}&\multirow{3}{0.9\textwidth}{$\displaystyle  -\frac{2 \varepsilon_{\mu\nu\rho\sigma}\varepsilon_{1i}^{\mu}p_{2}^{\nu}k_{1}^{\rho}k_{2}^{\sigma}  (2 m_2^2+s_1+s_2-3 s_3) s_3}{s_1^2+(s_2-s_3)^2-2 s_1 (s_2+s_3)}$}\vspace{15pt}\\[1pt]
      \multirow{3}{0.05\textwidth}{ $B_{7}^{2}$}&\multirow{3}{0.9\textwidth}{$\displaystyle \varepsilon_{\mu\nu\rho\sigma}\varepsilon_{1i}^{\mu}p_{2}^{\nu}k_{1}^{\rho}k_{2}^{\sigma}  \qty(-5+\frac{(m_1-m_2) (m_1+m_2)}{s_2}+\frac{2 (s_1^2+2 s_2 (s_2-2 s_3)+m_2^2 (-s_1+s_2+s_3)-s_1
      (3 s_2+s_3))}{s_1^2+(s_2-s_3)^2-2 s_1 (s_2+s_3)}) $}\vspace{15pt}\\[1pt]
      \multirow{4}{0.05\textwidth}{ $B_{7}^{3}$}&\multirow{4}{0.9\textwidth}{$\displaystyle  \frac{\varepsilon_{\mu\nu\rho\sigma}\varepsilon_{1i}^{\mu}p_{2}^{\nu}k_{1}^{\rho}k_{2}^{\sigma}  }{s_1 (s_1^2+(s_2-s_3)^2-2 s_1 (s_2+s_3))}(m_2^2 (s_1^2-(s_2-s_3)^2+4 s_1 s_3)+m_1^2 (s_1^2+(s_2-s_3)^2-2 s_1 (s_2+s_3))-s_1
      (s_1^2+3 s_2^2-8 s_2 s_3+5 s_3^2-2 s_1 (2 s_2+s_3)))$}\vspace{30pt}\\[1pt]

    \end{tabular}
  \end{ruledtabular}
\end{table}
\begin{table}[!ht]
  \centering
  \ContinuedFloat
  \caption{(Continued)}
  \begin{ruledtabular}
    \begin{tabular}{cc}
      Coefficient & Expression\\\hline
      \multirow{4}{0.05\textwidth}{ $C_{7}$}&\multirow{4}{0.9\textwidth}{$\displaystyle \frac{2 \varepsilon_{\mu\nu\rho\sigma}\varepsilon_{1i}^{\mu}p_{2}^{\nu}k_{1}^{\rho}k_{2}^{\sigma}  }{s_1^2+(s_2-s_3)^2-2 s_1 (s_2+s_3)}(2 m_2^4 s_3+s_3 (s_1^2+s_2^2-m_1^2 (s_1+s_2-3 s_3)-3 (s_1+s_2) s_3+2 s_3^2)-2 m_2^2
      ((s_1-s_2)^2+(m_1^2-3 (s_1+s_2)) s_3+3 s_3^2))$}\vspace{30pt}\\[1pt]
    \end{tabular}
  \end{ruledtabular}
\end{table}

\begin{table}[!ht]
  \centering
  \caption{Coefficients of scalar loop integrals in the decomposition of $i \mathcal{M}_{8}$}
  \begin{ruledtabular}   \label{tab:coeff8}
    \begin{tabular}{cc}
      Coefficient & Expression\\\hline
      \multirow{2}{0.05\textwidth}{ $A_{8}^{1}$}&\multirow{2}{0.9\textwidth}{$\displaystyle  -\frac{1}{2 s_2 (s_1^2+(s_2-s_3)^2-2 s_1 (s_2+s_3))} (2 \varepsilon_{1i}\cdot p_{2}\varepsilon_{1f}^{*}\cdot k_2 s_2+\varepsilon_{1i}\cdot p_{2}\varepsilon_{1f}^{*}\cdot p_2 (-s_1+s_2+s_3))$}\vspace{10pt}\\[1pt]
      \multirow{2}{0.05\textwidth}{ $A_{8}^{2}$}&\multirow{2}{0.9\textwidth}{$\displaystyle \frac{1}{2 s_2 (s_1^2+(s_2-s_3)^2-2 s_1 (s_2+s_3))} (2 \varepsilon_{1i}\cdot p_{2}\varepsilon_{1f}^{*}\cdot k_2 s_2+\varepsilon_{1i}\cdot p_{2}\varepsilon_{1f}^{*}\cdot p_2 (-s_1+s_2+s_3))$}\vspace{10pt}\\[1pt]
      \multirow{1}{0.05\textwidth}{ $A_{8}^{3}$}&\multirow{1}{0.9\textwidth}{$\setminus$}\\[1pt]
      \multirow{6}{0.05\textwidth}{ $B_{8}^{1}$}&\multirow{6}{0.9\textwidth}{$\displaystyle  \frac{1}{4
        (s_1^2+(s_2-s_3)^2-2 s_1 (s_2+s_3))^2}( 2 \varepsilon_{1i}\cdot p_{2}\varepsilon_{1f}^{*}\cdot k_2 ((-2 m_1^2+2 m_2^2+7 s_1-5 s_2) (s_1-s_2)^2+2 (-8 s_1^2+6 s_1 s_2+4 s_2^2-m_1^2 (s_1+s_2)+m_2^2
          (s_1+s_2)) s_3+(4 m_1^2-4 m_2^2+11 s_1-s_2) s_3^2-2 s_3^3)+2 \varepsilon_{1i}\cdot p_{2}\varepsilon_{1f}^{*}\cdot p_2 (8 (s_1-s_2)^3-(s_1-s_2)
          (-6 m_1^2+6 m_2^2+21 s_1+17 s_2) s_3+2 (3 m_1^2-3 m_2^2+6 s_1-5 s_2) s_3^2+s_3^3)+\varepsilon_{1i}\cdot \varepsilon_{1f}^{*} (4 (s_1-s_2)^2+2
      (m_1-m_2) (m_1+m_2) s_3-9 (s_1+s_2) s_3+5 s_3^2) (s_1^2+(s_2-s_3)^2-2 s_1 (s_2+s_3))) $}\vspace{45pt}\\[1pt]
      \multirow{7}{0.05\textwidth}{ $B_{8}^{2}$}&\multirow{7}{0.9\textwidth}{$\displaystyle  -\frac{1}{4 s_2 (s_1^2+(s_2-s_3)^2-2 s_1 (s_2+s_3))^2}(2 \varepsilon_{1i}\cdot p_{2}\varepsilon_{1f}^{*}\cdot p_2 (-(s_1-s_2)^2 (-m_1^2 (s_1-2 s_2)+m_2^2 (s_1-2 s_2)+3 s_2 (s_1+4 s_2))+(-m_1^2
            (3 s_1^2+2 s_1 s_2+3 s_2^2)+m_2^2 (3 s_1^2+2 s_1 s_2+3 s_2^2)+s_2 (11 s_1^2+32 s_1 s_2+29 s_2^2)) s_3+(3
          m_1^2 (s_1+2 s_2)-3 m_2^2 (s_1+2 s_2)-s_2 (13 s_1+22 s_2)) s_3^2+(-m_1^2+m_2^2+5 s_2) s_3^3)+s_2
          (2 \varepsilon_{1i}\cdot p_{2}\varepsilon_{1f}^{*}\cdot k_2 ((s_1-s_2) (-2 s_1^2-13 s_1 s_2+9 s_2^2+m_1^2 (s_1+5 s_2)-m_2^2 (s_1+5 s_2))+2
              (s_1 (-m_1^2+m_2^2+3 s_1)+2 (m_1^2-m_2^2+6 s_1) s_2+10 s_2^2) s_3+(m_1^2-m_2^2-6 s_1-13 s_2) s_3^2+2
            s_3^3)-\varepsilon_{1i}\cdot \varepsilon_{1f}^{*} ((s_1-s_2) (m_1^2-m_2^2-4 s_1+3 s_2)+(-m_1^2+m_2^2+8 s_1+7 s_2) s_3-4 s_3^2)
        (s_1^2+(s_2-s_3)^2-2 s_1 (s_2+s_3))))
      $}\vspace{55pt}\\[1pt]
      \multirow{6}{0.05\textwidth}{ $B_{8}^{3}$}&\multirow{6}{0.9\textwidth}{$\displaystyle  \frac{1}{2
        (s_1^2+(s_2-s_3)^2-2 s_1 (s_2+s_3))^2}(\varepsilon_{1i}\cdot p_{2}\varepsilon_{1f}^{*}\cdot p_{2} (-11 s_1^3+18 s_1^2 s_2-3 s_1 s_2^2-4 s_2^3-m_1^2 (s_1^2-2 s_1 (s_2-5 s_3)+(s_2-s_3)^2)+m_2^2
          (s_1^2-2 s_1 (s_2-5 s_3)+(s_2-s_3)^2)+32 s_1^2 s_3+28 s_1 s_2 s_3+12 s_2^2 s_3-25 s_1 s_3^2-12 s_2 s_3^2+4 s_3^3)+\varepsilon_{1i}\cdot p_{2}\varepsilon_{1f}^{*}\cdot k_{2}
          (-9 s_1^3+8 s_1^2 s_2+5 s_1 s_2^2-4 s_2^3+22 s_1^2 s_3+12 s_1 s_2 s_3+12 s_2^2 s_3-17 s_1 s_3^2-12 s_2 s_3^2+4 s_3^3+m_1^2 (5 s_1^2-(s_2-s_3)^2-4
        s_1 (s_2+s_3))+m_2^2 (-5 s_1^2+(s_2-s_3)^2+4 s_1 (s_2+s_3))))+\frac{1}{4
        (s_1^2+(s_2-s_3)^2-2 s_1 (s_2+s_3))} \varepsilon_{1i}\cdot \varepsilon_{1f}^{*} (-3 s_1^2+7 s_1 s_2-4 s_2^2+7 s_1 s_3+8 s_2 s_3-4
      s_3^2-m_1^2 (s_1-s_2+s_3)+m_2^2 (s_1-s_2+s_3))$}\vspace{45pt}\\[1pt]
      \multirow{9}{0.05\textwidth}{ $C_{8}$}&\multirow{9}{0.9\textwidth}{$\displaystyle  \frac{1}{2 (s_1^2+(s_2-s_3)^2-2 s_1 (s_2+s_3))^2}(-2 \varepsilon_{1i}\cdot p_{2}\varepsilon_{1f}^{*}\cdot k_2 ((s_1-s_2)^2 (m_1^4+m_2^4-s_1^2-2 m_1^2 (m_2^2+2 s_1-s_2)+m_2^2 (5 s_1-s_2)+6
            s_1 s_2-4 s_2^2)+(m_1^4 (s_1+s_2)+m_2^4 (s_1+s_2)+(5 s_1-7 s_2) (s_1^2-2 s_2^2)+m_2^2 (-11
            s_1^2+6 s_1 s_2+s_2^2)-2 m_1^2 (-4 s_1^2+4 s_1 s_2+2 s_2^2+m_2^2 (s_1+s_2))) s_3+(-2 m_1^4-2 m_2^4-9 s_1^2-11
          s_1 s_2-18 s_2^2+2 m_1^2 (2 m_2^2-2 s_1+s_2)+m_2^2 (7 s_1+s_2)) s_3^2+(-m_2^2+7 s_1+10 s_2) s_3^3-2 s_3^4)+2
          \varepsilon_{1i}\cdot p_{2}\varepsilon_{1f}^{*}\cdot p_2 ((4 m_1^2-3 m_2^2-4 s_2) (s_1-s_2)^3+(s_1-s_2) (3 m_1^4+3 m_2^4-s_1^2+6 s_1 s_2+14 s_2^2-2
            m_1^2 (3 m_2^2+6 s_1+5 s_2)+m_2^2 (11 s_1+7 s_2)) s_3+(3 m_1^4+3 m_2^4+4 s_1^2+m_1^2 (-6 m_2^2+6 s_1-8 s_2)+9
          s_1 s_2+18 s_2^2+m_2^2 (-7 s_1+5 s_2)) s_3^2-(-2 m_1^2+m_2^2+5 (s_1+2 s_2)) s_3^3+2 s_3^4)+\varepsilon_{1i}\cdot \varepsilon_{1f}^{*}
          (-(s_1-s_2)^2 (-2 m_1^2+3 m_2^2+2 (s_1+s_2))+(m_1^4+m_2^4+8 s_1^2+s_1 s_2+8 s_2^2+7 m_2^2 (s_1+s_2)-m_1^2
          (2 m_2^2+5 (s_1+s_2))) s_3+(3 m_1^2-4 m_2^2-10 (s_1+s_2)) s_3^2+4 s_3^3) (s_1^2+(s_2-s_3)^2-2
      s_1 (s_2+s_3)))  $}\vspace{80pt}\\[1pt]
    \end{tabular}
  \end{ruledtabular}
\end{table}
\begin{table}[!ht]
  \centering
  \caption{Coefficients of scalar loop integrals in the decomposition of $i \mathcal{M}_{9}$}
  \begin{ruledtabular}   \label{tab:coeff9}
    \begin{tabular}{cc}
      Coefficient & Expression\\\hline
      \multirow{1}{0.05\textwidth}{ $A_{9}^{1}$}&\multirow{1}{0.9\textwidth}{$\displaystyle 0$}\\[1pt]
      \multirow{1}{0.05\textwidth}{ $A_{9}^{2}$}&\multirow{1}{0.9\textwidth}{$\displaystyle 0$}\\[1pt]
      \multirow{1}{0.05\textwidth}{ $A_{9}^{3}$}&\multirow{1}{0.9\textwidth}{$\setminus$}\\[1pt]
      \multirow{2}{0.05\textwidth}{ $B_{9}^{1}$}&\multirow{2}{0.9\textwidth}{$\displaystyle \frac{2 \varepsilon_{1f}^{*}\cdot p_{2} s_3+\varepsilon_{1f}^{*}\cdot k_{2} (-s_1+s_2+s_3)}{s_1^2+(s_2-s_3)^2-2 s_1 (s_2+s_3)} $}\vspace{10pt}\\[1pt]
      \multirow{2}{0.05\textwidth}{ $B_{9}^{2}$}&\multirow{2}{0.9\textwidth}{$\displaystyle -\frac{2 \varepsilon_{1f}^{*}\cdot k_{2} s_2+\varepsilon_{1f}^{*}\cdot p_{2} (-s_1+s_2+s_3)}{s_1^2+(s_2-s_3)^2-2 s_1 (s_2+s_3)} $}\vspace{10pt}\\[1pt]

    \end{tabular}
  \end{ruledtabular}
\end{table}
\begin{table}[!ht]
  \centering
  \ContinuedFloat
  \caption{(Continued)}
  \begin{ruledtabular}
    \begin{tabular}{cc}
      Coefficient & Expression\\\hline
      \multirow{2}{0.05\textwidth}{ $B_{9}^{3}$}&\multirow{2}{0.9\textwidth}{$\displaystyle \frac{\varepsilon_{1f}^{*}\cdot k_{2} (s_1+s_2-s_3)-\varepsilon_{1f}^{*}\cdot p_{2} (s_1-s_2+s_3)}{s_1^2+(s_2-s_3)^2-2 s_1 (s_2+s_3)}$}\vspace{10pt}\\[1pt]
      \multirow{4}{0.05\textwidth}{ $C_{9}$}&\multirow{4}{0.9\textwidth}{$\displaystyle \frac{1}{s_1^2+(s_2-s_3)^2-2 s_1 (s_2+s_3)}(\varepsilon_{1f}^{*}\cdot k_{2} (m_1^2-m_2^2-s_2) (-s_1+s_2)+\varepsilon_{1f}^{*}\cdot k_{2} (m_1^2-m_2^2+s_2) s_3+\varepsilon_{1f}^{*}\cdot p_{2} ((s_1-s_2)^2+2
      (m_1-m_2) (m_1+m_2) s_3-3 (s_1+s_2) s_3+2 s_3^2))$}\vspace{30pt}\\[1pt]
    \end{tabular}
  \end{ruledtabular}
\end{table}

\begin{table}[!ht]
  \centering
  \caption{Coefficients of scalar loop integrals in the decomposition of $i \mathcal{M}_{10}$}
  \begin{ruledtabular}   \label{tab:coeff10}
    \begin{tabular}{cc}
      Coefficient & Expression\\\hline
      \multirow{1}{0.05\textwidth}{ $A_{10}^{1}$}&\multirow{1}{0.9\textwidth}{$\displaystyle 0$}\\[1pt]
      \multirow{1}{0.05\textwidth}{ $A_{10}^{2}$}&\multirow{1}{0.9\textwidth}{$\displaystyle 0$}\\[1pt]
      \multirow{1}{0.05\textwidth}{ $A_{10}^{3}$}&\multirow{1}{0.9\textwidth}{$\setminus$}\\[1pt]
      \multirow{1}{0.05\textwidth}{ $B_{10}^{1}$}&\multirow{1}{0.9\textwidth}{$\displaystyle -1$}\\[1pt]
      \multirow{1}{0.05\textwidth}{ $B_{10}^{2}$}&\multirow{1}{0.9\textwidth}{$\displaystyle 1$}\\[1pt]
      \multirow{1}{0.05\textwidth}{ $B_{10}^{3}$}&\multirow{1}{0.9\textwidth}{$\displaystyle 1$}\\[1pt]
      \multirow{1}{0.05\textwidth}{ $C_{10}$}&\multirow{1}{0.9\textwidth}{$ -m_1^2+2 m_2^2+s_1+s_2-2 s_3 $}\\[1pt]
    \end{tabular}
  \end{ruledtabular}
\end{table}

\begin{table}[!ht]
  \centering
  \caption{Coefficients of scalar loop integrals in the decomposition of $i \mathcal{M}_{11}$}
  \begin{ruledtabular}   \label{tab:coeff11}
    \begin{tabular}{cc}
      Coefficient & Expression\\\hline
      \multirow{2}{0.05\textwidth}{ $A_{11}^{1}$}&\multirow{2}{0.9\textwidth}{$\displaystyle -\frac{(s_1+s_2) (2 M_f^2+s_1+s_2-s_3-2 s_{23})}{2 s_1 s_2}$}\vspace{10pt}\\[1pt]
      \multirow{2}{0.05\textwidth}{ $A_{11}^{2}$}&\multirow{2}{0.9\textwidth}{$\displaystyle \frac{(s_1+s_2) (2 M_f^2+s_1+s_2-s_3-2 s_{23})}{2 s_1 s_2}$}\vspace{10pt}\\[1pt]
      \multirow{1}{0.05\textwidth}{ $A_{11}^{3}$}&\multirow{1}{0.9\textwidth}{$\setminus$}\\[1pt]
      \multirow{3}{0.05\textwidth}{ $B_{11}^{1}$}&\multirow{3}{0.9\textwidth}{$\displaystyle -\frac{2 (-m_1^2+2 m_2^2+s_1+s_2-2 s_3) s_3 (2 M_f^2+s_1+s_2-s_3-2 s_{23})}{s_1^2+(s_2-s_3)^2-2 s_1 (s_2+s_3)}$}\vspace{20pt}\\[1pt]
      \multirow{4}{0.05\textwidth}{ $B_{11}^{2}$}&\multirow{4}{0.9\textwidth}{$\displaystyle \frac{1}{2 s_2 (s_1^2+(s_2-s_3)^2-2 s_1 (s_2+s_3))}(-((s_1-s_2) (-m_1^2 (s_1+s_2)+s_2 (3 s_1+s_2)+m_2^2 (s_1+3 s_2)))+2
        (4 s_1 s_2-m_1^2 (s_1+2 s_2)+m_2^2 (s_1+3 s_2)) s_3+(m_1^2-m_2^2-5 s_2) s_3^2) (2 M_f^2+s_1+s_2-s_3-2
      s_{23})$}\vspace{25pt}\\[1pt]
      \multirow{4}{0.05\textwidth}{ $B_{11}^{3}$}&\multirow{4}{0.9\textwidth}{$\displaystyle \frac{1}{2
        s_1 (s_1^2+(s_2-s_3)^2-2 s_1 (s_2+s_3))}((s_1-s_2) (m_2^2 (3 s_1+s_2)+s_1 (s_1+3 s_2))+2 (4 s_1 s_2+m_2^2 (3 s_1+s_2))
      s_3-(m_2^2+5 s_1) s_3^2+m_1^2 (-s_1^2+(s_2-s_3)^2-4 s_1 s_3)) (2 M_f^2+s_1+s_2-s_3-2 s_{23})$}\vspace{25pt}\\[1pt]
      \multirow{3}{0.05\textwidth}{ $C_{11}$}&\multirow{3}{0.9\textwidth}{$\displaystyle \frac{(-m_1^2+2 m_2^2+s_1+s_2-2 s_3) (-2 m_1^2+2 m_2^2+s_1+s_2-s_3) s_3 (2 M_f^2+s_1+s_2-s_3-2 s_{23})}{s_1^2+(s_2-s_3)^2-2
      s_1 (s_2+s_3)} $}\vspace{20pt}\\[1pt]
    \end{tabular}
  \end{ruledtabular}
\end{table}

\begin{table}[!ht]
  \centering
  \caption{Coefficients of scalar loop integrals in the decomposition of $i \mathcal{M}_{12}$}
  \begin{ruledtabular}   \label{tab:coeff12}
    \begin{tabular}{cc}
      Coefficient & Expression\\\hline
      \multirow{1}{0.05\textwidth}{ $A_{12}^{1}$}&\multirow{1}{0.9\textwidth}{$\displaystyle 0$}\\[1pt]
      \multirow{1}{0.05\textwidth}{ $A_{12}^{2}$}&\multirow{1}{0.9\textwidth}{$\displaystyle 0$}\\[1pt]
      \multirow{1}{0.05\textwidth}{ $A_{12}^{3}$}&\multirow{1}{0.9\textwidth}{$\setminus$}\\[1pt]
      \multirow{3}{0.05\textwidth}{ $B_{12}^{1}$}&\multirow{3}{0.9\textwidth}{$\displaystyle \frac{(-2 m_1^2+2 m_2^2+s_1+s_2-s_3) s_3^2 (2 M_f^2+s_1+s_2-s_3-2 s_{23})}{2 (s_1^2+(s_2-s_3)^2-2 s_1 (s_2+s_3))}$}\vspace{20pt}\\[1pt]
    \end{tabular}
  \end{ruledtabular}
\end{table}
\begin{table}[!ht]
  \centering
  \ContinuedFloat
  \caption{(Continued)}
  \begin{ruledtabular}
    \begin{tabular}{cc}
      Coefficient & Expression\\\hline
      \multirow{3}{0.05\textwidth}{ $B_{12}^{2}$}&\multirow{3}{0.9\textwidth}{$\displaystyle\frac{s_3 ((s_1-s_2) (-m_1^2+m_2^2+s_2)+(m_1^2-m_2^2+s_2) s_3) (2 M_f^2+s_1+s_2-s_3-2 s_{23})}{2
      (s_1^2+(s_2-s_3)^2-2 s_1 (s_2+s_3))}$}\vspace{20pt}\\[1pt]
      \multirow{3}{0.05\textwidth}{ $B_{12}^{3}$}&\multirow{3}{0.9\textwidth}{$\displaystyle -\frac{s_3 ((-m_1^2+m_2^2+s_1) (s_1-s_2)-(m_1^2-m_2^2+s_1) s_3) (2 M_f^2+s_1+s_2-s_3-2 s_{23})}{2
      (s_1^2+(s_2-s_3)^2-2 s_1 (s_2+s_3))} $}\vspace{20pt}\\[1pt]
      \multirow{4}{0.05\textwidth}{ $C_{12}$}&\multirow{4}{0.9\textwidth}{$\displaystyle -\frac{1}{s_1^2+(s_2-s_3)^2-2 s_1 (s_2+s_3)} s_3 (m_2^4 s_3+(m_1^4+s_1 s_2-m_1^2 (s_1+s_2-s_3)) s_3+m_2^2 ((s_1-s_2)^2-(2 m_1^2+s_1+s_2)
      s_3)) (2 M_f^2+s_1+s_2-s_3-2 s_{23})$}\vspace{30pt}\\[1pt]
    \end{tabular}
  \end{ruledtabular}
\end{table}

\bibliography{0.Dynamics_TS.bib}
\end{document}